\documentclass[a4paper,11pt]{article}

\usepackage{jheppub} 

\usepackage{preprintcover}  
\PreprintCoverPaperTitle{Measurement of isolated-photon pair production in 
$pp$ collisions at $\sqrt{s} = 7 \tev$ with the ATLAS detector}  
\PreprintIdNumber{CERN-PH-EP-2012-300}  
\PreprintCoverAbstract{The ATLAS experiment at the LHC
  has measured the production cross section of
  events with two isolated photons in the final state, in
  proton-proton collisions at $\sqrt{s} = 7~\tev$.
  The full data set collected in 2011, corresponding to an
  integrated luminosity of $4.9~\ifb$, is used. 
  The amount of background, from hadronic jets and isolated electrons,
  is estimated with data-driven techniques and subtracted.
  The total cross section, for two isolated photons with transverse
  energies above 25 GeV and 22 GeV respectively, in the acceptance
  of the electromagnetic calorimeter ($|\eta|<1.37$ and $1.52<|\eta|<2.37$) 
  and with an angular separation $\Delta R>0.4$, is $44.0^{+3.2}_{-4.2}$~pb.
  The differential cross sections as a function of the \diphoton
  invariant mass, transverse momentum, azimuthal separation,
  and cosine of the polar angle of the largest transverse energy photon in the
  Collins--Soper \diphoton rest frame are also measured.
  The results are compared to the prediction of leading-order parton-shower and
  next-to-leading-order and next-to-next-to-leading-order parton-level generators.}  
\PreprintJournalName{JHEP}  

\usepackage[T1]{fontenc} 

\usepackage{atlasphysics} 

\def\diphoton{di-photon }

\newcommand{\ie} {{\it i.e.}}

\newcommand{\Et}{\ensuremath{E_{\mathrm T}}}
\newcommand{\EtOne}{\ensuremath{E_{\mathrm T,1}}}
\newcommand{\EtTwo}{\ensuremath{E_{\mathrm T,2}}}

\newcommand{\gamjet} {$\gamma$-jet}

\newcommand{\wgg} {\ensuremath{N_{\mathrm{\evtgamgam}}}}
\newcommand{\wgj} {\ensuremath{N_{\mathrm{\evtgamjet}}}}
\newcommand{\wjg} {\ensuremath{N_{\mathrm{\evtjetgam}}}}
\newcommand{\wjj} {\ensuremath{N_{\mathrm{\evtjetjet}}}}

\newcommand{\mgg}{\ensuremath{m_\evtgamgam}}
\newcommand{\ptgg}{\ensuremath{p_{{\rm T},\evtgamgam}}}
\newcommand{\dphigg}{\ensuremath{\Delta\phi_\evtgamgam}}
\newcommand{\cosThetagg}{\ensuremath{\cos\theta^*_\evtgamgam}}

\newcommand{\Etisol} {\ensuremath{\ET^{\mathrm{iso}}}}
\newcommand{\EtisolOne} {\ensuremath{E_\mathrm{T,1}^{\mathrm{iso}}}}
\newcommand{\EtisolTwo} {\ensuremath{E_\mathrm{T,2}^{\mathrm{iso}}}}
\newcommand{\EtisolI}   {\ensuremath{E_{\mathrm{T},i}^{\mathrm{iso}}}}

\newcommand{\tight}{\ensuremath{\mathbf{T}}}
\newcommand{\isol}{\ensuremath{\mathbf{I}}}
\newcommand{\nontight}{\ensuremath{\mathbf{\tilde{T}}}}
\newcommand{\nonisol}{\ensuremath{\mathbf{\tilde{I}}}}

\newcommand{\Tight}{tight}
\newcommand{\nonTight}{non-tight}

\newcommand{\evtgamgam}{\ensuremath{{\gamma\gamma}}}
\newcommand{\evtgamjet}{\ensuremath{{\gamma\mathrm{j}}}}
\newcommand{\evtjetgam}{\ensuremath{{\mathrm{j}\gamma}}}
\newcommand{\evtjetjet}{\ensuremath{{\mathrm{jj}}}}

\newcommand{\el}{\ensuremath{e}}
\newcommand{\ph}{\ensuremath{\gamma}}

\newcommand{\fph}{\ensuremath{f_{\el\rightarrow\ph}}}
\newcommand{\fel}{\ensuremath{f_{\ph\rightarrow\el}}}

\newcommand{\ngel}  {\ensuremath{N_{\ph\el}}}
\newcommand{\nelel} {\ensuremath{N_{\el\el}}}

\newcommand{\ngelZ}  {\ensuremath{N^Z_{\ph\el}}}
\newcommand{\nelelZ} {\ensuremath{N^Z_{\el\el}}}
\newcommand{\nelelelZ}  {\ensuremath{N^Z_{\el\el\el}}}
\newcommand{\ngelelZ}   {\ensuremath{N^Z_{\ph\el\el}}}
\newcommand{\nggsig} {\ensuremath{N^{\rm sig}_{\mathrm{\evtgamgam}}}}

\newcommand{\pythia}{\textsc{Pythia}}
\newcommand{\sherpa}{\textsc{Sherpa}}
\newcommand{\diphox}{\textsc{Diphox}}
\newcommand{\alpgen}{\textsc{Alpgen}}
\newcommand{\gammatwomc}{\textsc{gamma2mc}}
\newcommand{\twognnlo}{\textsc{2$\gamma$NNLO}}

\title{\boldmath Measurement of isolated-photon pair production in 
$pp$ collisions at $\sqrt{s} = 7 \tev$ with the ATLAS detector}

\author{G. Aad \textit{et al.} (full author list given at the end of the article in Appendix)}
\collaboration{ATLAS Collaboration}

\abstract{The ATLAS experiment at the LHC
  has measured the production cross section of
  events with two isolated photons in the final state, in
  proton-proton collisions at $\sqrt{s} = 7~\tev$.
  The full data set collected in 2011, corresponding to an
  integrated luminosity of $4.9~\ifb$, is used. 
  The amount of background, from hadronic jets and isolated electrons,
  is estimated with data-driven techniques and subtracted.
  The total cross section, for two isolated photons with transverse
  energies above 25 GeV and 22 GeV respectively, in the acceptance
  of the electromagnetic calorimeter ($|\eta|<1.37$ and $1.52<|\eta|<2.37$) 
  and with an angular separation $\Delta R>0.4$, is $44.0^{+3.2}_{-4.2}$~pb.
  The differential cross sections as a function of the \diphoton
  invariant mass, transverse momentum, azimuthal separation,
  and cosine of the polar angle of the largest transverse energy photon in the
  Collins--Soper \diphoton rest frame are also measured.
  The results are compared to the prediction of leading-order parton-shower and
  next-to-leading-order and next-to-next-to-leading-order parton-level generators.
}

\begin{document} 

\maketitle
\flushbottom

\section{Introduction}
\label{s:Introduction}
The measurement at the LHC of the production cross section, in $pp$ collisions,
of two isolated photons not originating from hadronic decays,
$pp\to \gamma\gamma+X$, provides a tool to probe
perturbative Quantum Chromodynamics (QCD) predictions and to
understand the irreducible background to new physics processes 
involving photons in the final state. These processes include 
Higgs boson decays to photon pairs ($H\rightarrow\evtgamgam$)
or graviton decays predicted in some Universal Extra-Dimension models \cite{ued_rs,ued_add}.

Recent cross section measurements for \diphoton production at hadron
colliders were performed by the D\O~\cite{D0_diphoton} and
CDF~\cite{CDF_diphoton} collaborations at the $\sqrt{s}=1.96$~TeV
Tevatron $p\bar{p}$ collider, and by ATLAS~\cite{Aad:2011mh} and
CMS~\cite{CMS_diphoton} using $\sqrt{s}=7$~TeV $pp$ collisions recorded
at the LHC in 2010. 

In this paper, the production cross section of two isolated photons with transverse 
energies (\ET) above 25 GeV and 22 GeV respectively, in the
acceptance of the ATLAS electromagnetic calorimeter 
($|\eta|< 1.37$ and $1.52 <|\eta| < 2.37$) and with
an angular separation $\Delta R > 0.4$, is measured.
The results are obtained using the data collected by the ATLAS
experiment in 2011, which corresponds to an integrated luminosity\footnote{The 3.9\% uncertainty 
in the integrated luminosity for the complete 2011 data set 
is based on the calibration described in refs. \cite{ATLASluminosity,ATLAS-OLDLUMI} including an additional 
uncertainty for the extrapolation to the later data-taking period with higher 
instantaneous luminosity.} of $(4.9\pm 0.2)~\ifb$, thus increasing the sample size by more 
than a factor of 100 compared to the previous measurement.
The transverse energy thresholds for the two photons
are higher than in the previous measurement (16 GeV).

The integrated \diphoton production cross section is measured,
as well as the differential cross sections as a function of four kinematic
variables: the \diphoton invariant mass (\mgg), the \diphoton transverse
momentum (\ptgg), the azimuthal\footnote{ATLAS uses a right-handed coordinate system with its
origin at the nominal interaction point (IP) in the centre of the detector 
and the $z$-axis coinciding with the axis of the
beam pipe. The $x$-axis points from the IP to the centre of
the LHC ring, and the $y$-axis points upward. Cylindrical coordinates
($r$,$\phi$) are used in the transverse plane, $\phi$ being the
azimuthal angle around the beam pipe. The pseudorapidity
is defined in terms of the polar angle $\theta$ as $\eta = -\ln \tan(\theta/2)$.}
 separation ~between the photons
in the laboratory frame (\dphigg), and the cosine of the polar
angle of the highest $\ET$ photon in 
the Collins--Soper \diphoton rest frame (\cosThetagg)~\cite{CollinsSoper}. 
The first distribution is of obvious interest for resonance searches;
the second and the third provide important information in the study of
higher-order QCD perturbative effects and fragmentation,
especially in some specific regions such as the small \dphigg\ limit;
the fourth can be used to investigate the spin of \diphoton
resonances. For this purpose, the Collins-Soper rest frame is preferred  to other 
frame definitions because of its robustness with respect to initial state radiation.
The results are compared to the predictions from: parton-shower Monte
Carlo generators, \pythia~\cite{pythia} and \sherpa~\cite{sherpa}; 
parton-level calculations with next-to-leading-order (NLO) QCD 
corrections using the \diphox~\cite{diphox} program complemented
by \gammatwomc~\cite{gamma2MC}; and at next-to-next-to-leading-order (NNLO),
using \twognnlo~\cite{2gNNLO}. 
The contribution from the \diphoton decays of the particle recently discovered by 
ATLAS~\cite{ATLAS_Higgs} and CMS~\cite{CMS_Higgs} in the search
for the Standard Model Higgs boson is not included in the theoretical
calculations. It is expected to contribute around $1\%$ of the signal
in the 120$<\mgg<$130~GeV interval, and negligibly elsewhere.

\section{The ATLAS detector}
\label{s:Atlas}
ATLAS~\cite{ATLAS_detector} is a multipurpose detector with a forward-backward 
symmetric cylindrical geometry and nearly 4\pi\ coverage in solid angle. 
The most relevant subdetectors for the present 
analysis are the inner tracking detector (ID) and the calorimeters.
 
The ID consists of a silicon pixel detector and a silicon microstrip
detector covering the pseudorapidity
range $|\eta|<2.5$,
and a straw tube transition radiation tracker covering $|\eta|<2.0$. 
It is immersed in a 2 T magnetic field provided by a superconducting
solenoid.
The ID allows efficient reconstruction of converted photons if
the conversion occurs at a radius of up to $\approx 0.80$ m.

The electromagnetic calorimeter (ECAL) is a lead/liquid-argon (LAr)
sampling calo\-ri\-me\-ter providing coverage for $|\eta|< 3.2$.
It consists of a barrel section ($|\eta| < 1.475$) and two
end-caps ($1.375<|\eta|< 3.2$). The central region ($|\eta|<2.5$)
is segmented into three longitudinal layers.
The first (inner) layer, covering $|\eta| < 1.4$ in the barrel and 
$1.5<|\eta|< 2.4$ in the end-caps, has high granularity in the $\eta$
direction (between 0.003 and 0.006 depending on $\eta$),
sufficient to provide event-by-event discrimination
between single-photon showers and two overlapping
showers from a $\pi^0$ decay. 
The second layer, which collects most of the energy deposited 
in the calorimeter by the photon shower, has a cell 
granularity of 0.025$\times$0.025 in $\eta\times\phi$. 
The third layer is used to correct high energy showers for leakage beyond
the ECAL. In front of the electromagnetic
calorimeter a thin presampler layer, covering the
pseudorapidity interval $|\eta|< 1.8$, is used to correct for
energy loss before the ECAL.

The hadronic calorimeter (HCAL), surrounding the
ECAL, consists of an iron/scin\-til\-la\-tor tile calorimeter
in the range $|\eta|< 1.7$, and two copper/LAr calorimeters
spanning $1.5 <|\eta|< 3.2$. The ECAL and HCAL acceptance is extended
by two LAr forward calorimeters (using copper and tungsten
as absorbers) up to $|\eta|< 4.9$.

Di-photon events are recorded using a three-level trigger system.
The first level, implemented in
hardware, is based on towers defined with a coarser  granularity ($0.1 \times 0.1$ in
$\eta\times\phi$) than that of the ECAL. 
They are used to search for electromagnetic deposits in $\eta\times\phi$ regions of $2\times1$
and $1\times2$ towers, 
within a fixed window of size $2\times 2$ and with a 
transverse energy above a programmable threshold. 
The second- and third-level triggers are implemented in
software and exploit the full granularity and energy calibration of
the calorimeter to refine the first-level trigger selection.

\section{Data and Monte Carlo samples}
\label{s:Data}
The data set analysed consists of the 7~TeV proton-proton collisions
recorded by the ATLAS detector in 2011.
Only events where the beam conditions are stable and the 
trigger system, the tracking devices, and the calorimeters are 
operational, are considered.

Monte Carlo (MC) samples are produced using various generators as
described below. 
Particle interactions with the detector material
and the detector response are simulated  with \textsc{Geant4}~\cite{geant}.
The events are reconstructed with the same algorithms
used for collision data.
More details of the event generation and simulation 
infrastructure are provided in ref.~\cite{ATLASsimulation}.

Simulated \diphoton events are generated with both \pythia~6.4.21
and \sherpa~1.3.1. \pythia\ uses the modified leading-order 
MRST2007~\cite{MRST2007} parton distribution functions (PDFs)
while \sherpa\ uses the CTEQ6L1~\cite{cteq} PDFs.
The \pythia\ event-generator parameters are set according to the 
ATLAS AMBT2~\cite{AMBT2} tune, while the \sherpa\ parameters
are the default ones of the \sherpa~1.3.1 distribution.
Photons originating from the hard scattering and quark bremsstrahlung are included in the analysis. 
The MC \diphoton signal is generated with a photon $\Et$ threshold of 20~GeV;
one million events are produced both with \pythia\ and \sherpa.
They are used to model the transverse isolation energy (see section~\ref{s:Selection}) distribution
of signal photons, to compute the reconstruction efficiency and
to study the systematic uncertainties on the reconstructed quantities. 
Background $\gamma$-jet events are generated using \alpgen~\cite{alpgen} with
the CTEQ6L1 PDF set.


\section{Event selection}
\label{s:Selection}
Events are collected using a \diphoton trigger with a nominal
transverse energy threshold of 20 GeV for both photon candidates.
The photon trigger objects are required to pass a selection
based on shower shape variables computed from the energy 
deposits in the second layer of the electromagnetic calorimeter 
and in the hadronic calorimeter.
The requirements are looser than the photon identification criteria applied in
the offline selection.
In order to reduce non-collision backgrounds, events are required to
have a reconstructed primary vertex with at least three associated tracks
and consistent with the average beam spot position.
The signal inefficiency of this requirement is negligible.

Photons are reconstructed from electromagnetic energy clusters in the
calorimeter and tracking information provided by the ID as
described in ref.~\cite{ATLAS-INCLPHOTON}. Photons reconstructed near
regions of the calorimeter affected by read-out or high-voltage
failures are not considered.\footnote{This requirement leads to 
a typical loss of 0.8\% to 1.4\% on the photon reconstruction
efficiency, depending on the data-taking period.} 
The cluster energies are corrected 
using an in-situ calibration based on the $Z$ boson
mass peak~\cite{electronperf_2011}, and the determination of
the pseudorapidities
is optimized using the technique described in ref.~\cite{ATLAS_Higgs}.
In order to benefit from the fine segmentation of the first
layer of the electromagnetic calorimeter to discriminate between 
genuine prompt photons and fake photons within jets,
the photon candidate pseudorapidity 
must satisfy $|\eta|<1.37$ or $1.52<|\eta|<2.37$.
We retain photon candidates passing loose identification requirements,
based on the same shower shape variables -- computed with better granularity and resolution --
and the same thresholds used at trigger level.
The highest-\Et\ (``leading'') and second highest-\Et\ (``subleading'') 
photons within the acceptance and satisfying the loose identification criteria 
are required to have $\EtOne >25$ GeV and $\EtTwo >22$ GeV, respectively.
The fraction of events where the two selected photon 
candidates are not matched to the photon trigger objects is negligible.
The angular separation between the two photons, $\Delta
R=\sqrt{(\Delta\eta)^2+(\Delta\phi)^2}$, is required to be
larger than 0.4, in order to avoid one photon candidate
depositing significant energy in the isolation cone
of the other, as defined below. 

Two further criteria are used to define the signal and background
control regions. 
Firstly the \Tight\ photon selection~\cite{ATLAS-INCLPHOTON}
(abbreviated as \tight\ in the following) is
designed to reject hadronic jet background, by imposing requirements
on nine discriminating variables computed from the energy leaking 
into the HCAL and the lateral and longitudinal 
shower development in the ECAL.
Secondly the transverse isolation energy \Etisol\ is computed from the sum of
the positive-energy topological clusters with reconstructed barycentres
inside a cone of radius $\Delta R=0.4$ around the photon candidate.
The algorithm for constructing topological clusters suppresses noise by keeping
only those cells with a significant energy deposit and their
neighbouring cells. 
The cells within $0.125 \times 0.175$ in $\eta \times \phi$ around the 
photon are excluded from the calculation of \Etisol.
The mean value of the small leakage of the photon energy
from this region into the isolation cone, evaluated as a function of the
photon transverse energy, is subtracted from the measured
value of \Etisol~ (meaning that \Etisol~ can be negative).
The typical size of this correction is a few percent of the
photon transverse energy. 
The measured value of \Etisol~is further corrected by
subtracting the estimated contributions from the underlying event 
and additional $pp$ interactions.
This correction is computed on an event-by-event basis, 
by calculating the transverse energy density from
low-transverse-momentum~jets, 
as suggested in refs.~\cite{Cacciari:area,Cacciari:UE}.
The median transverse energy densities of the jets 
in two $\eta$ regions, $|\eta|<1.5$ and $1.5<|\eta|<3.0$, 
are computed separately, and the one for the region 
containing the photon candidate pseudorapidity is
multiplied by the total area of all topological clusters 
used in the calculation of the isolation variable in order to
estimate the correction.
Signal photons are required to pass the \Tight\
selection (``tight photons'') and the isolation requirement \isol, $-4<\Etisol<4$~GeV. 
A total of 165~767 pairs of \Tight, isolated photons are selected.
The fraction of events in which an additional photon pair passes all
the selection criteria, except for the requirement on the two photons
being the leading and subleading $\Et$ candidates, is less than 1 per 
100~000. 
The \nonTight\ (\nontight) photon candidates are defined as those
failing the \Tight\ criteria for at least one of the
shower-shape variables that are computed from the energy
deposits in a few cells of the first layer of the electromagnetic calorimeter adjacent 
to the cluster barycentre.
Photon candidates with $4<\Etisol<8$~GeV are considered non-isolated (\nonisol).

\section{Signal yield extraction}
\label{s:yield_extraction}
After the selection, the main background is due primarily to
\gamjet~ and secondarily to di-jet (\evtjetjet) final states,
collectively called ``jet background'' in the following.
Two methods, the two-dimensional sidebands and the two-dimensional fit,
already exploited in ref.~\cite{Aad:2011mh}, are used 
to perform an in-situ statistical subtraction of the jet background
from the selected photon candidate pairs, as described in section~\ref{ss:jet_subtraction}.

After the jet background contribution is subtracted, a small
residual background contamination arises from events where isolated 
electrons are misidentified as photons. 
This contribution is estimated as described in section~\ref{ss:el_subtraction}.

\subsection{Jet background subtraction}
\label{ss:jet_subtraction}

Both the two-dimensional sidebands and the two-dimensional fit methods
use the photon transverse isolation energy and the
\Tight\ identification criteria to discriminate prompt photons from
jets. 
They rely on the fact that the correlations between the isolation and the 
\Tight\ criteria in background events are small, and that the signal
contamination in the \nonTight\ or non-isolated control regions is
low.

The two-dimensional sidebands method counts the numbers of photon candidate
pairs where each of the candidates passes or fails the \Tight\ 
and the isolation criteria. 
Four categories are defined for each photon, resulting in 16 categories of events.
The inputs to the method are the numbers
of events in the categories and the 
signal efficiencies of the \Tight\ and isolation requirements. The
correlation between these two requirements is assumed to be negligible
for background events. The method allows the
simultaneous extraction of the numbers of true \diphoton signal, \evtgamjet, \evtjetgam
\footnote{Here and in the following, \evtgamjet\ (\evtjetgam) denotes the events
where the leading (subleading) candidate is a true photon, and the other candidate a true jet.}
and \evtjetjet\ background events, and the \Tight\ and isolation 
efficiencies for fake photon candidates from jets (``fake rates'').
The expected number of events in each category is written
as a function of the parameters (yields, efficiencies, fake rates and
correlation factors) and the system of 16 equations is solved with a
$\chi^2$ minimization procedure. 
This method is an extension of the one used in our previous di-photon
analysis~\cite{Aad:2011mh}. It allows the extraction of different isolation fake
rates for jets in \evtjetgam\ or \evtjetjet\ events as well as a
correlation factor for the isolation of jet pairs. 
 
The two-dimensional fit method consists of an extended maximum 
likelihood template fit to the two-dimensional distribution of the transverse
isolation energies \EtisolOne\ and \EtisolTwo\ of the
two photon candidates in events belonging to the \tight-\tight\ sample,
\ie\ where both photons satisfy the \Tight\ identification
criteria. The fit is performed in the isolation range $-4 < \EtisolI < 8~{\rm GeV}$ $(i=1,2)$.
The correlations between the transverse isolation energies of the two
candidates in di-photon, \evtgamjet, and \evtjetgam\ events are 
found to be negligible in MC samples, and the products of two one-dimensional templates for
\EtisolOne\ and \EtisolTwo\ are used for each of the three event species. 
For the \evtjetjet\ component, large correlations 
are observed in data, and a two-dimensional template is used.
The two-dimensional fit is described in detail in our previous
paper~\cite{Aad:2011mh}. There are two differences between the 
present and previous analyses: the use now of binned
distributions instead of smooth parametric functions for the photon and jet
templates, and the correction for signal leakage in the background
templates, as described below.

The transverse isolation energy distributions of the signal photons 
and the corresponding efficiencies of the signal requirement $-4<\Etisol<4$~GeV
are obtained from the \sherpa\ \diphoton sample, separately for the
leading and the subleading candidates.
In the two-dimensional fit method, the templates are shifted 
by +160 and +120 MeV respectively 
in order to maximize the likelihood, as
determined from a scan as a function of the shifts. These values
are also used to compute the signal efficiencies of the isolation 
requirement needed in the two-dimensional sidebands method.
Shifts of similar size between ATLAS data and MC simulation have been observed
in the transverse isolation energy distribution, computed
with the same technique (based on topological clusters inside a cone
of radius 0.4), of electron control samples selected from $Z\to ee$ decays
with a tag-and-probe method. 
The \Etisol\ distributions of prompt photons in \evtgamjet\ and
\evtjetgam\ events are assumed to be identical to that of prompt
photons in \diphoton events, as found in simulated samples.
The \Tight\ identification efficiencies for prompt photons, needed in the two-dimensional sidebands method and
 in the final cross section measurement, are estimated using the same \diphoton
MC sample. The shower shape variables are corrected for the observed
differences between data and simulation in photon-enriched control
samples. Residual differences between the efficiencies in the simulation
and in data are corrected using scale factors determined from control
samples of photons from radiative $Z$ boson decays, electrons selected with 
a tag-and-probe technique from $Z\to ee$ decays, and photon-enriched
control samples of known photon purity~\cite{ATLAS-CONF-2012-123}.
After applying these corrections, the photon identification efficiency 
in the simulation is estimated to reproduce the efficiency in data to
within 2\%.
For the two-dimensional fit, the transverse isolation energy template
of the leading (subleading) jet in \evtjetgam\ (\evtgamjet) events is
extracted directly from data where one candidate passes the
\nonTight\ and the other passes both the tight identification and
isolation (\tight\isol) requirements. 
For \evtjetjet\ events, the two-dimensional template is obtained from data
in which the two candidates are required to be \nonTight. The correlation 
is found to be about 8\%. 
The jet background templates are corrected for signal leakage in
the control samples, estimated from the \sherpa\ sample.

Figure~\ref{fig:TF_global} shows the projections of the
two-dimensional fit to the transverse isolation energies of the leading
and subleading photon candidates. 
The yields for each of the four components extracted with the 
two-dimensional sidebands method and the two-dimensional fit
are given in table~\ref{tab:yields}. The 
\diphoton purity is around 68\% and the \diphoton yields
agree within 1.5\% between the two methods. 

\begin{figure}[hbtp]
  \centering
  \includegraphics[width=0.48\columnwidth]{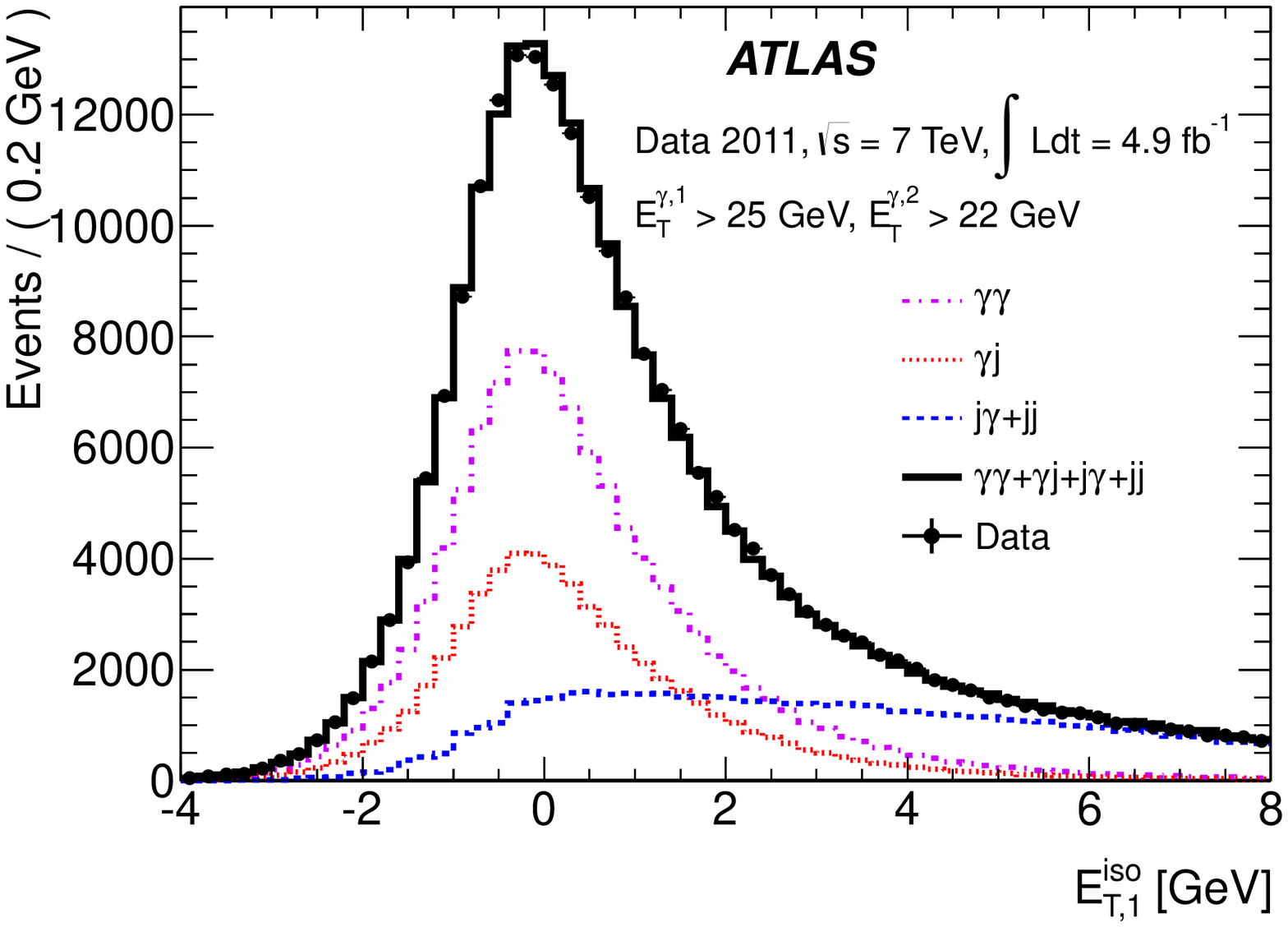}
  \includegraphics[width=0.48\columnwidth]{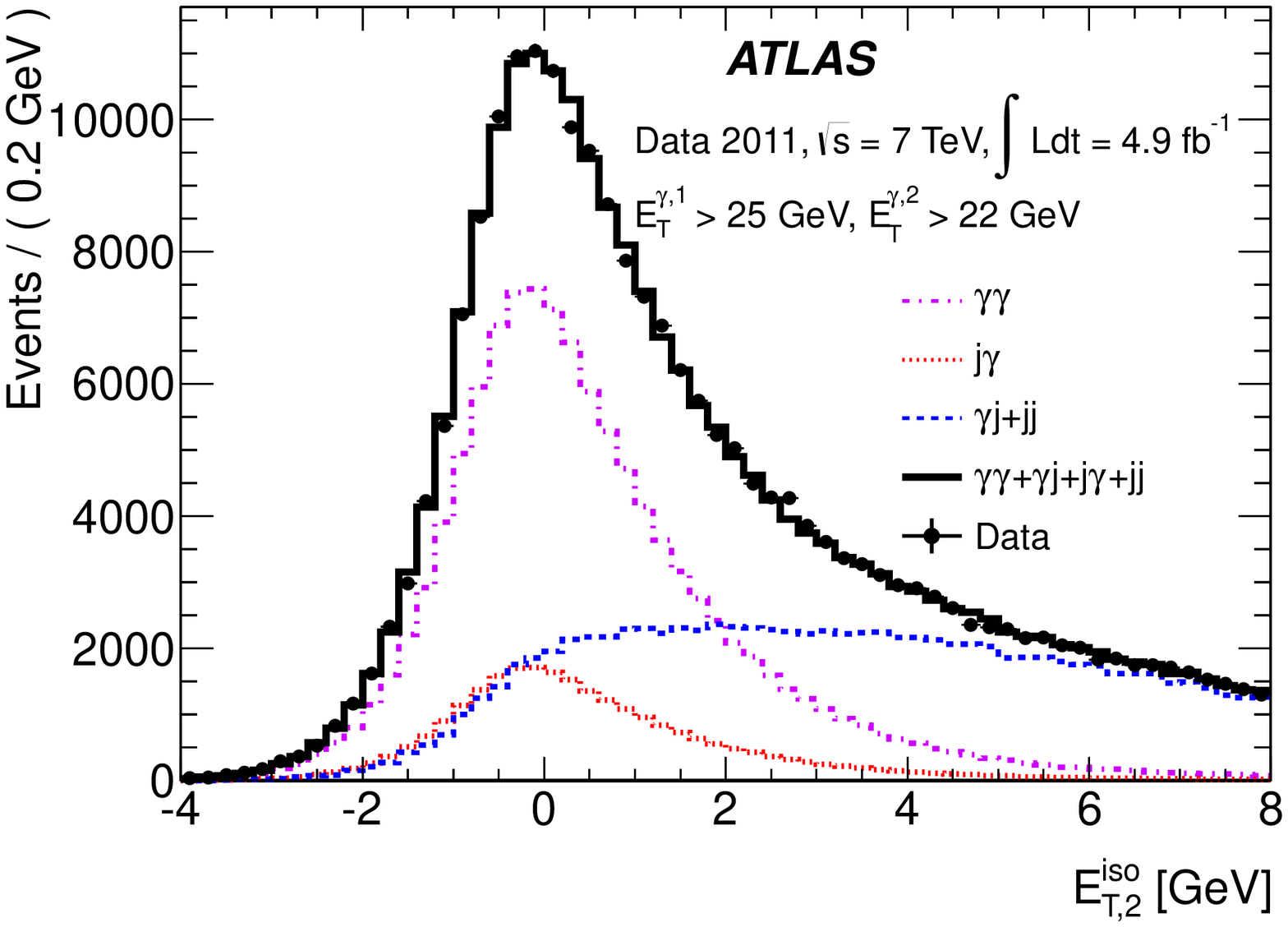}
  \caption{
    Projections of the two-dimensional fit to the transverse isolation energies of the two photon candidates:
    leading photon (left) and sub-leading photon (right). The photon templates from \sherpa\ are 
    shifted by +160 MeV (+120 MeV) for the leading (subleading) photon.  
    Solid circles represent the observed data. The (black) solid line
    is the fit result, the (violet) dash-dotted curve shows the \evtgamgam\ component. The (red) dotted line shows in the left (right) figure
    the contribution from \evtgamjet\ (\evtjetgam) events.
    In both figures, the (blue) dashed line represents a broad background component in the photon candidates' sample: for the leading
    candidate this is due to \evtjetgam\ and \evtjetjet\ final states, whereas for the sub-leading candidate it comes from \evtgamjet\
    and \evtjetjet\ final states. 
  }
  \label{fig:TF_global}
\end{figure}

 \begin{table}[!htb]
   \begin{center}
     \begin{tabular}{lrccrcc}
       \hline\hline
       Yield    & \multicolumn{3}{c}{two-dimensional sidebands results} & \multicolumn{3}{c}{two-dimensional fit results} \\ \hline
 \wgg &  113\,200 & $\pm 600$ (stat.) & $_{-8000}^{+5000}$ (syst.) & 111\,700 & $\pm 500$ (stat.)     & $_{-7600}^{+4500}$ (syst.) \\ 
 \wgj &   31\,500 & $\pm 400$ (stat.) & $_{-3100}^{+3900}$ (syst.) &  31\,500 & $\pm 300$ (stat.)     & $_{-3600}^{+4800}$ (syst.) \\ 
 \wjg &   13\,000 & $\pm 300$ (stat.) & $_{-\phantom{0}800}^{+2500}$ (syst.) &  13\,900 & $^{+300}_{-200}$ (stat.) & $_{-2100}^{+3400}$ (syst.) \\
 \wjj &    8\,100 & $\pm 100$ (stat.) & $_{-1400}^{+1900}$ (syst.) &   8\,300 & $\pm 100$ (stat.)     & $_{-2100}^{+\phantom{0}300}$ (syst.) \\
       \hline\hline
     \end{tabular}
   \end{center}
   \caption{
     Total yields for two candidates satisfying the \Tight\ identification and the isolation requirement $-4 < \Etisol < 4$ GeV. 
     Both statistical and total systematic uncertainties are listed.
     \label{tab:yields}
   }
 \end{table}

To obtain the differential signal yields as a function of the \diphoton
kinematic variables, such as \mgg, \ptgg, \dphigg\ and \cosThetagg, the
above methods are applied in each bin of the variables. Figure
\ref{fig:TF_statOnly} shows the differential spectra of the 
signal and background components obtained with the two-dimensional fit. 
In some regions of the \diphoton spectra, discrepancies with the
two-dimensional sidebands results are larger than those observed for the 
integrated yield. 
The results from the two-dimensional fit are used to extract the nominal cross sections,
while differences between the results obtained with the two methods 
are included in the final systematic uncertainty.

\begin{figure}[!hbtp]
  \centering
  \includegraphics[width=0.48\columnwidth]{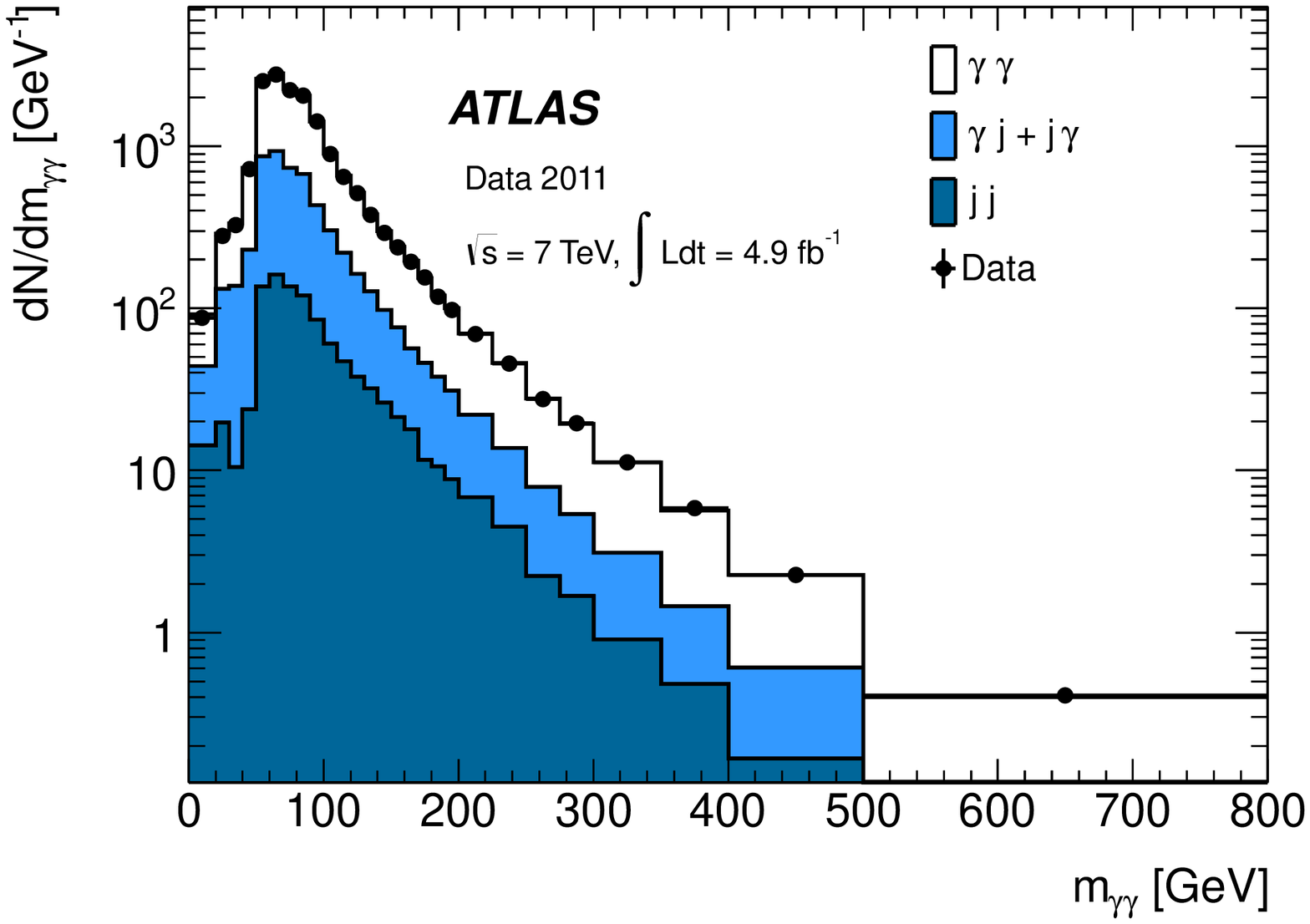}
  \includegraphics[width=0.48\columnwidth]{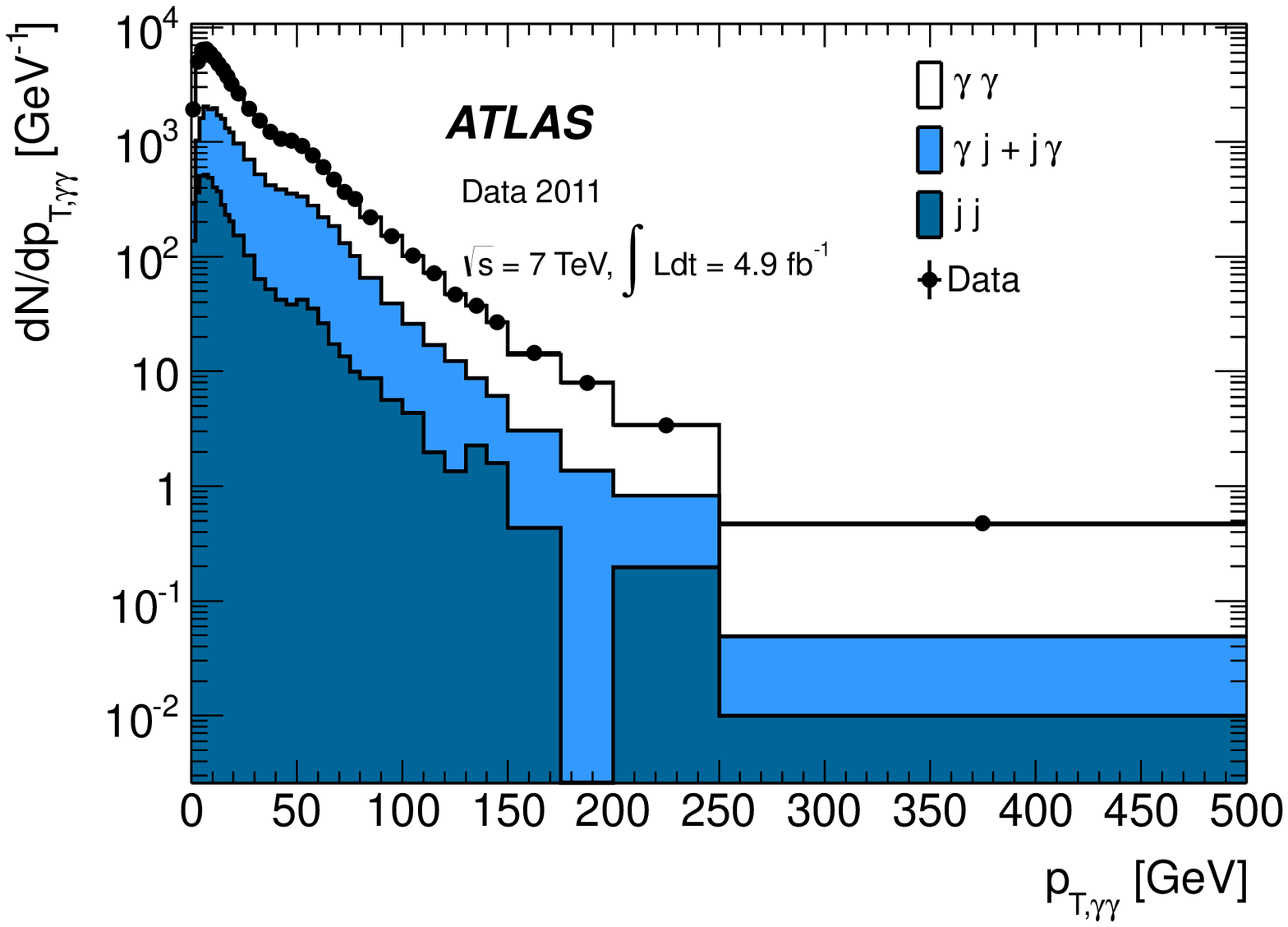}
  \includegraphics[width=0.48\columnwidth]{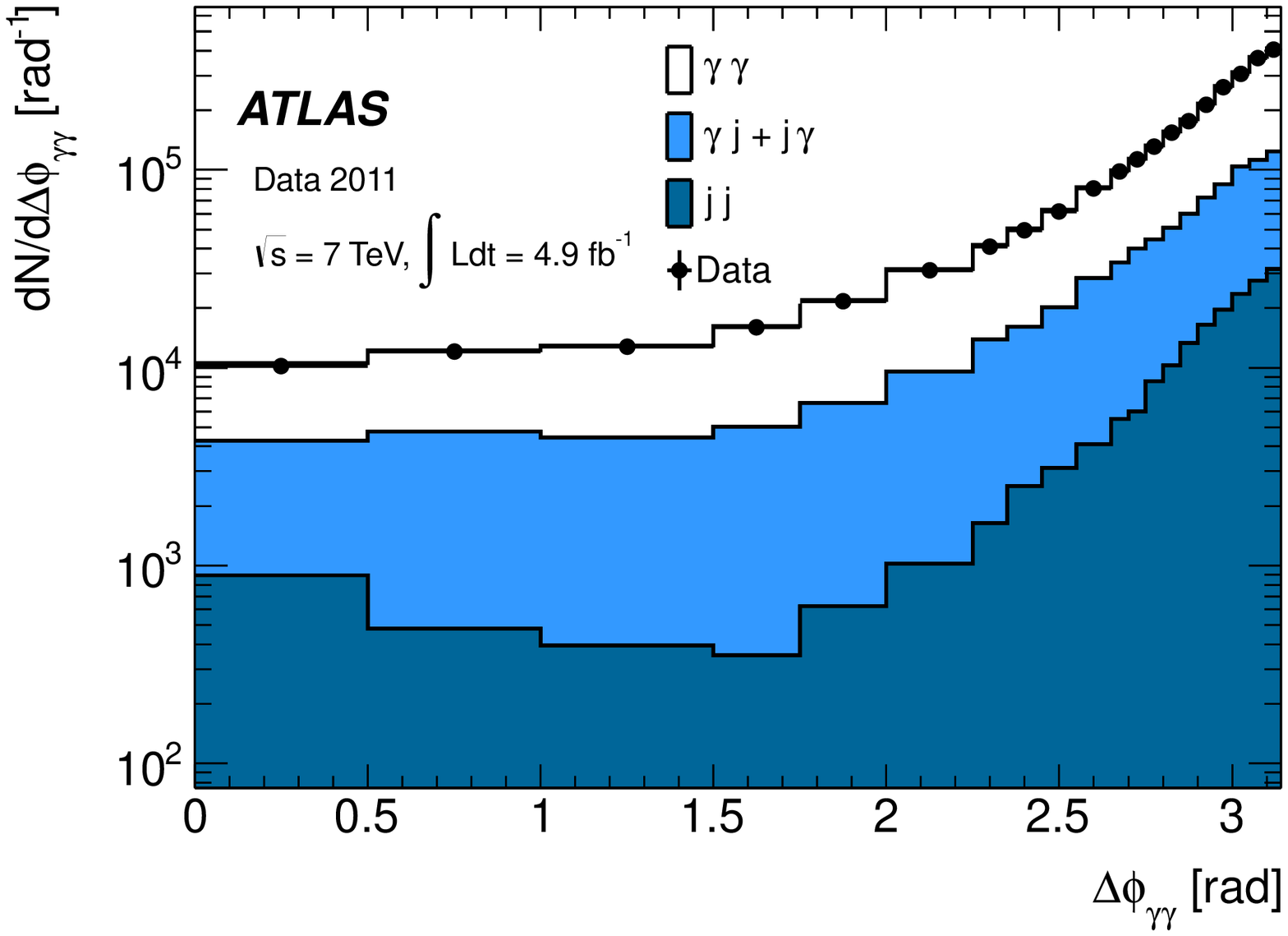}
  \includegraphics[width=0.48\columnwidth]{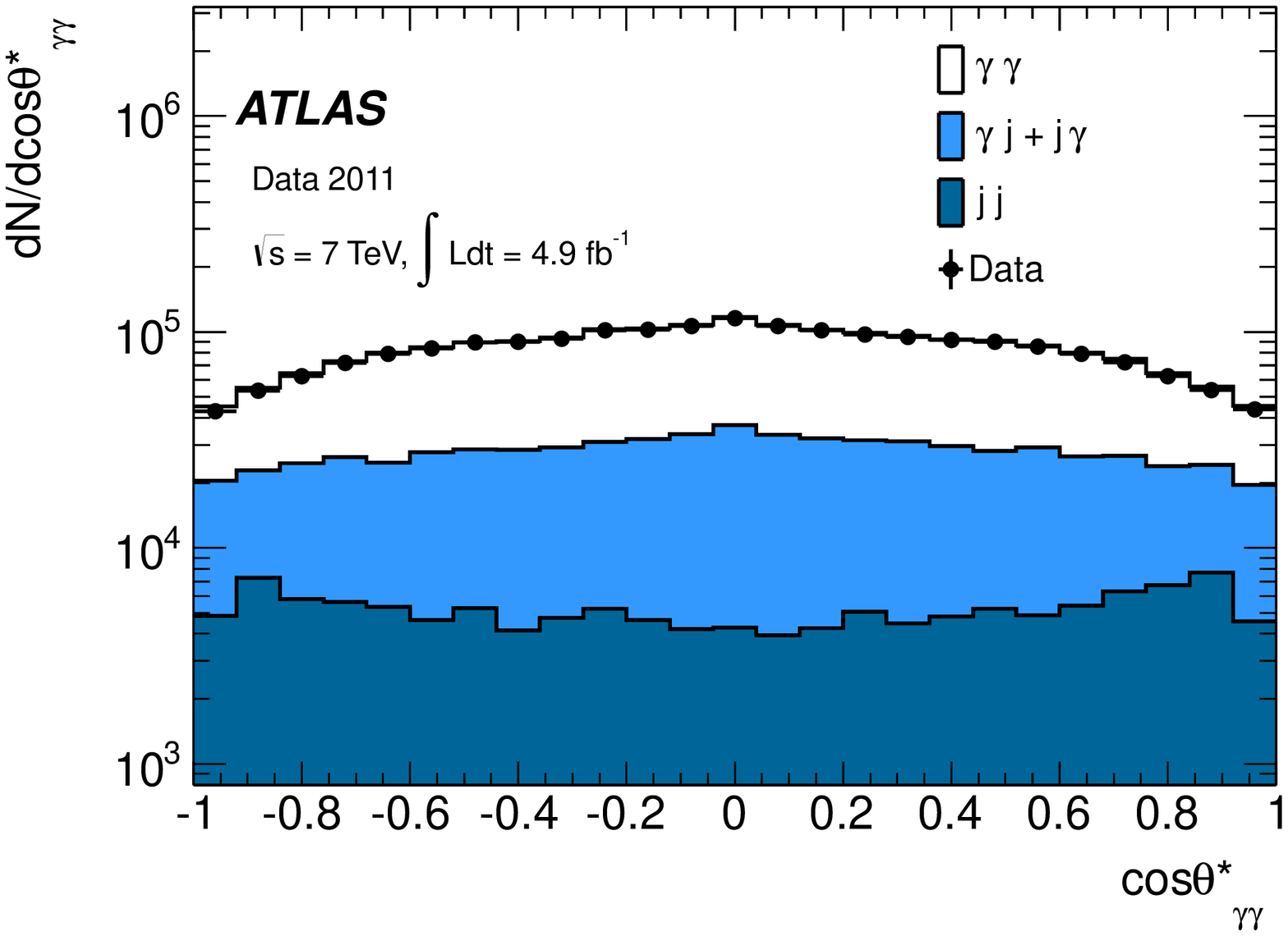}
  \caption{
    Differential spectra in data (solid circles) and from the
    two-dimensional fit, for the $\gamma\gamma$ (hollow histogram),
    $\gamma$j$+$j$\gamma$ (light solid histogram), and jj (dark solid histogram)
    contributions. The spectra are shown for the following \diphoton\
    variables: \mgg~(top left), \ptgg~(top right),
    \dphigg~(bottom left), \cosThetagg\ (bottom right).   
  }
  \label{fig:TF_statOnly}
\end{figure}

Several sources of systematic uncertainty on the signal yield, 
estimated after the jet background subtraction, are considered.
The dominant uncertainty originates from the choice of the
background control regions and accounts for both the uncertainty 
on the background transverse isolation energy distribution and 
its correlation with the identification criteria. 
It is first estimated by varying the number
of relaxed criteria in the \nonTight\ definition. 
For the 
integrated \diphoton yield, the effect is found to be $_{-6}^{+3}$\%.
In some bins of the \mgg\ and \ptgg\ spectra where the
size of the control samples is small, neighbouring bins are
grouped together to extract the jet background templates. 
Since the background transverse isolation energies depend mildly
on these kinematic variables, a systematic uncertainty is evaluated 
by repeating the yield extraction with jet templates from the adjacent 
groups of neighbouring bins. The uncertainty on the estimated 
signal yield is at most $\pm 9$\%.

In the nominal result, the photon isolation templates are taken from the
\sherpa\ \diphoton sample.
A systematic uncertainty is evaluated by using alternative templates
from the \pythia\ \diphoton sample, and from data. 
The data--driven template for the leading (subleading) photon
is obtained by selecting events where the requirement $\Etisol<$ 8~GeV
is removed for the leading (subleading) photon candidate, and normalizing the leading
(subleading) photon isolation distribution in
\nontight-\tight\isol~(\tight\isol-\nontight) events, where the leading
(subleading) candidate fails the tight identification while the other
candidate passes tight identification and isolation criteria, to the
isolation distribution of leading (subleading) candidates in
\tight-\tight~events in the $7<\Etisol<17$ GeV region. The
difference between the two distributions is used as an estimate of the
photon distribution.
The \pythia\ \diphoton sample exhibits higher tails than \sherpa\ at large values of 
$\Etisol$. The data-driven template, on the other hand, is characterized by
smaller tails than the \sherpa\ template, since it is obtained by 
assuming that the isolation region above 7 GeV is fully populated by background. 
The corresponding uncertainty on the signal yield is estimated to be $_{-3}^{+2}$\%
of the 
integrated \diphoton yield. It is rather uniform as a function of \mgg, \ptgg, \dphigg~ and \cosThetagg\ and always
below 4\%, except at very low \mgg\ where it reaches $\pm$5\%. 
The photon isolation template is, to a large extent, independent
of the variables under study. Repeating the background subtraction 
procedure using photon isolation templates extracted in bins of the
\diphoton variable under study leads to variations of the estimated
signal yield within $_{-4}^{+2}$\%.

Other systematic effects have been considered, and found to be
smaller than those previously discussed. The bias created by neglecting the dependence of the identification and isolation efficiencies
on $\eta$ and \Et~is estimated to be of +0.02\% and -0.3\% respectively. The effect
of assuming identical templates for photons in \diphoton and in
\gamjet\ events is 
evaluated by using instead templates from \alpgen\ \gamjet\ samples for photons in the \evtgamjet\ and \evtjetgam\ components. 
The uncertainty on the shifts applied to the MC photon templates ($\pm$10~MeV for the leading photons and $\pm$5~MeV for 
the subleading ones, as determined from the scan) is propagated to the 
\diphoton yields. 
The impact of the identification efficiencies on the 
signal leakage correction is estimated by neglecting in the 
simulation the correction factors nominally applied to the shower shape variables
to account for the observed differences between data and MC simulation. These effects produce
systematic uncertainties of at most 0.5\% on the differential spectra. Finally, no significant
effect is observed due to the imperfect modelling of the material in front of the calorimeters.

\subsection{Electron background subtraction}
\label{ss:el_subtraction}

Isolated electrons from $W$ or $Z$ boson decays can be misidentified as photons,
since the two particles ($e$ and $\gamma$) generate
similar electromagnetic showers in the ECAL.
Usually a track is reconstructed in the inner detector pointing to the
electron  ECAL cluster, thus isolated electrons misidentified as
photons are mostly classified as converted candidates.
Pairs of misidentified, isolated electrons and positrons ($ee$) from 
processes such as Drell--Yan, $Z\rightarrow\el\el$,
$WW\rightarrow\el\nu\el\nu$, or of photons and $e^\pm$ from diboson
production ($\ph W\rightarrow\ph\el\nu$, $\ph Z\rightarrow\ph\el\el$),
provide a background that cannot be distinguished from the \diphoton signal  
based on the photon identification and isolation variables and must therefore
be estimated in a separate way.
The same procedure exploited in ref.~\cite{Aad:2011mh}, based on the number of 
\ph\el\ (\ngel) and \el\el\ (\nelel) events observed in data,
is used to estimate their contributions to the \diphoton yield
\wgg\ after jet background subtraction.

For a given bin $i$ of the variable $X$ ($X=$ \mgg, \ptgg, \dphigg, or \cosThetagg), the signal component \nggsig\
 in the \wgg\ sample can be evaluated:
\begin{equation}
  \nggsig = \frac{\wgg-\left[ \fph \ngel  - (\fph)^2 \nelel  \right]}{(1-\fph\fel)^2}
  \label{eq:elbkg-mgg-from-N}
\end{equation}
The fake rates \fph\ and \fel\ are measured using $Z$ boson decays in data. 
$Z\rightarrow\el\el$ decays are used to estimate $\fph$ as $\ngelZ/(2\nelelZ)$,
where \ngelZ\ and \nelelZ\ are the numbers of $\gamma e$ and $ee$ pairs with invariant mass
within $1.5~\sigma$ of the $Z$ boson mass. 
$Z\rightarrow\ph\el\el$ decays are similarly used to estimate $\fel$ 
as $\nelelelZ/\ngelelZ$.
The numbers of continuum background events are estimated from the
sidebands of the $ee$, $\gamma e$, $eee$ or $\gamma ee$ invariant mass
distributions ($51-61\,\mathrm{GeV}$ and $121-131\,\mathrm{GeV}$),
and subtracted from \nelelZ, \ngelZ, \nelelelZ\ and 
\ngelelZ, respectively.
Electrons must satisfy identification criteria based on their 
shower shape in the electromagnetic calorimeter, quality criteria for 
the associated track in the ID, and an isolation requirement
$\Etisol<4~\rm{GeV}$.
The measured fake rates, including statistical and systematic uncertainties,
are $\fph=0.062_{-0.010}^{+0.040}$ and $\fel=0.038_{-0.007}^{+0.024}$, where
the systematic uncertainty is dominated by the dependence on the
transverse energy of the candidate photon.
Other sources include the uncertainties on \nelelZ, \ngelZ, \nelelelZ\
and \ngelelZ\ which are evaluated by changing the definition of the $Z$ boson mass window to 
$\pm 2\sigma$ and $\pm 1\sigma$, and shifting the sidebands by $\pm 5\,\mathrm{GeV}$. 
The fraction of electron background as a function of \mgg,
\ptgg, \dphigg, and \cosThetagg\ is shown in figure~\ref{fig:electron}. 
The enhancements at $\mgg\approx m_Z$, low \ptgg\ and $\dphigg\approx\pi$ are
due to the large $Z$ boson production cross section.

\begin{figure}[!htb]
 \begin{center}
   \includegraphics[width=0.48\columnwidth]{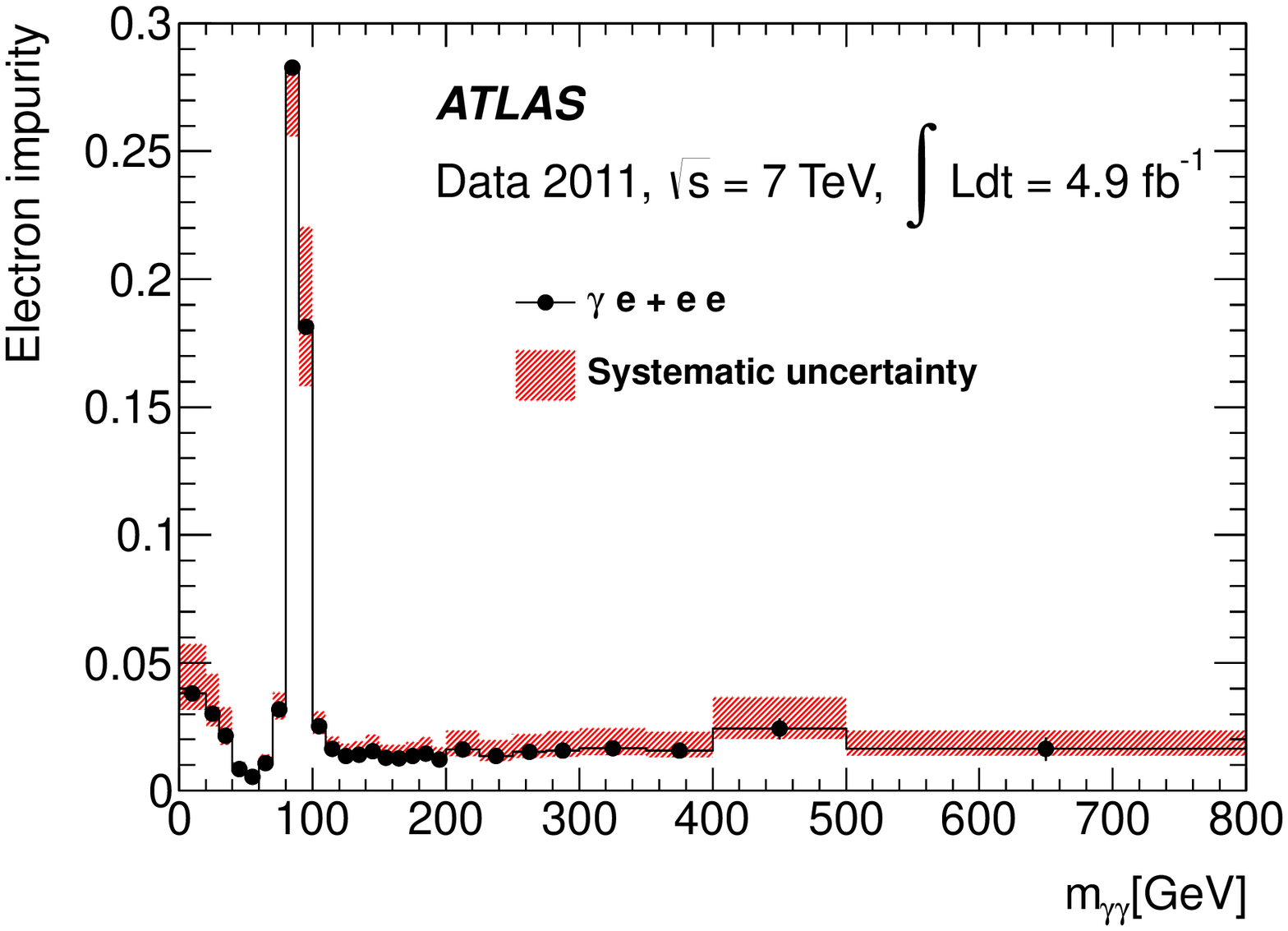}
   \includegraphics[width=0.48\columnwidth]{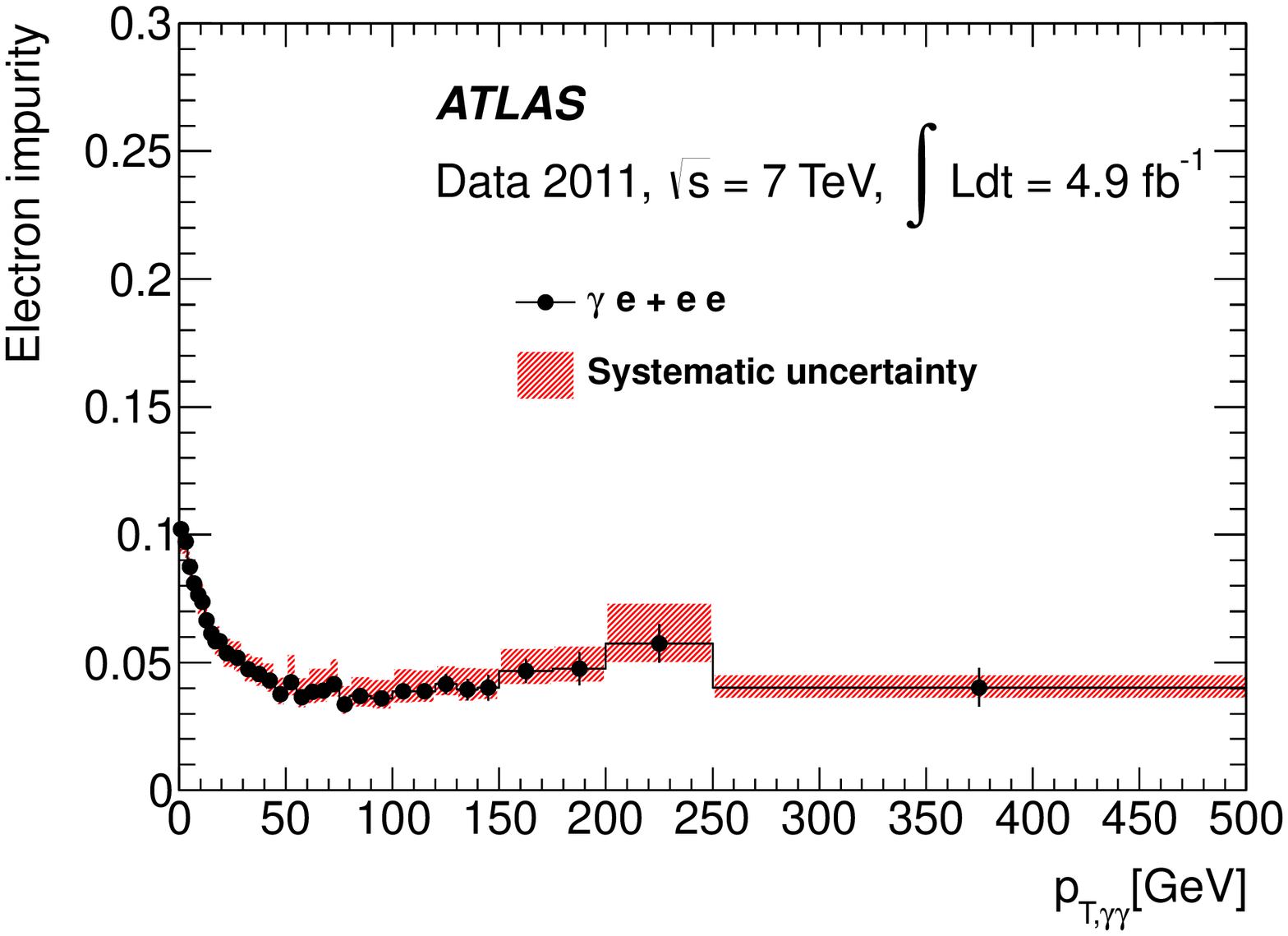}
   \includegraphics[width=0.48\columnwidth]{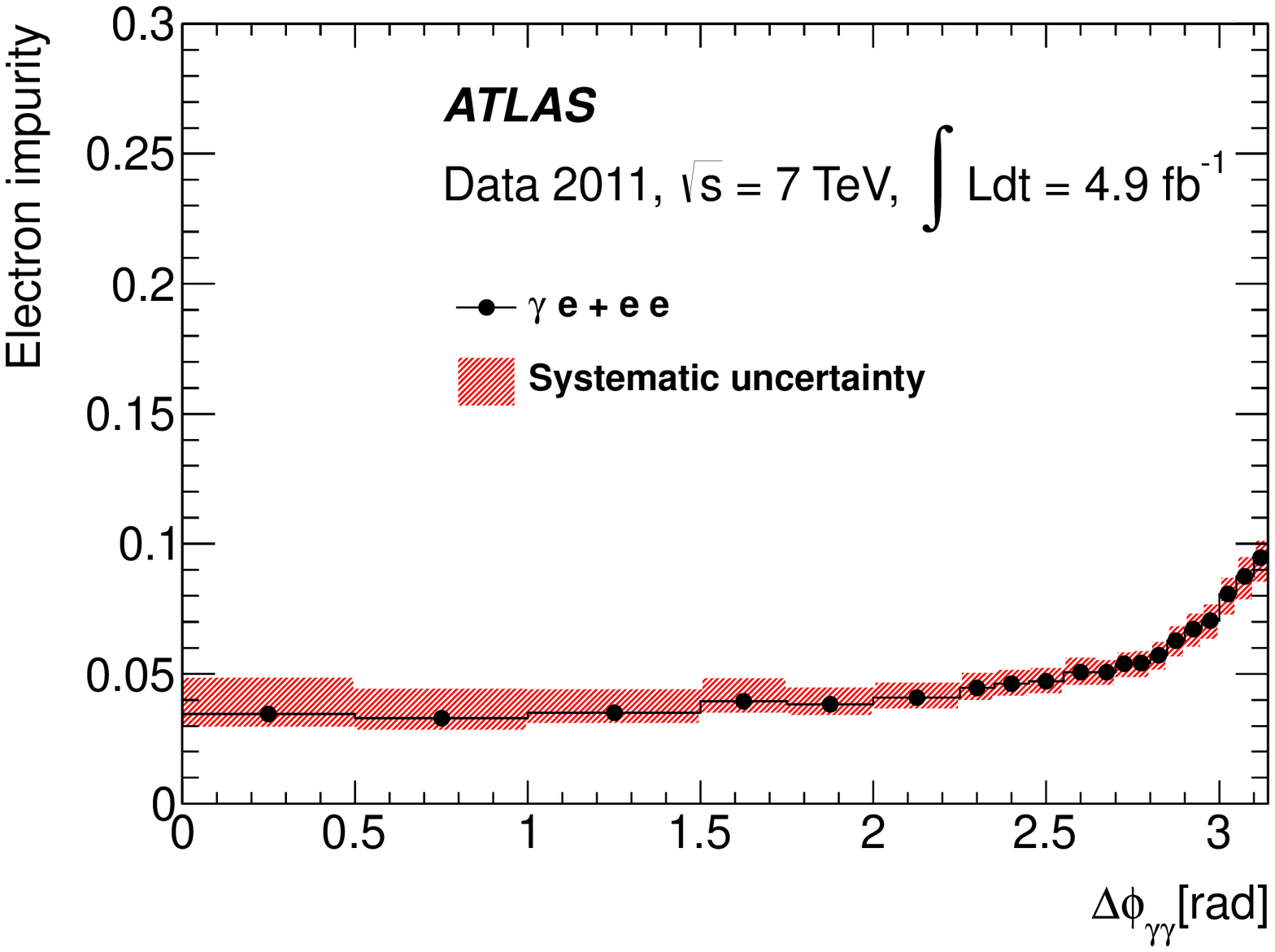}
   \includegraphics[width=0.48\columnwidth]{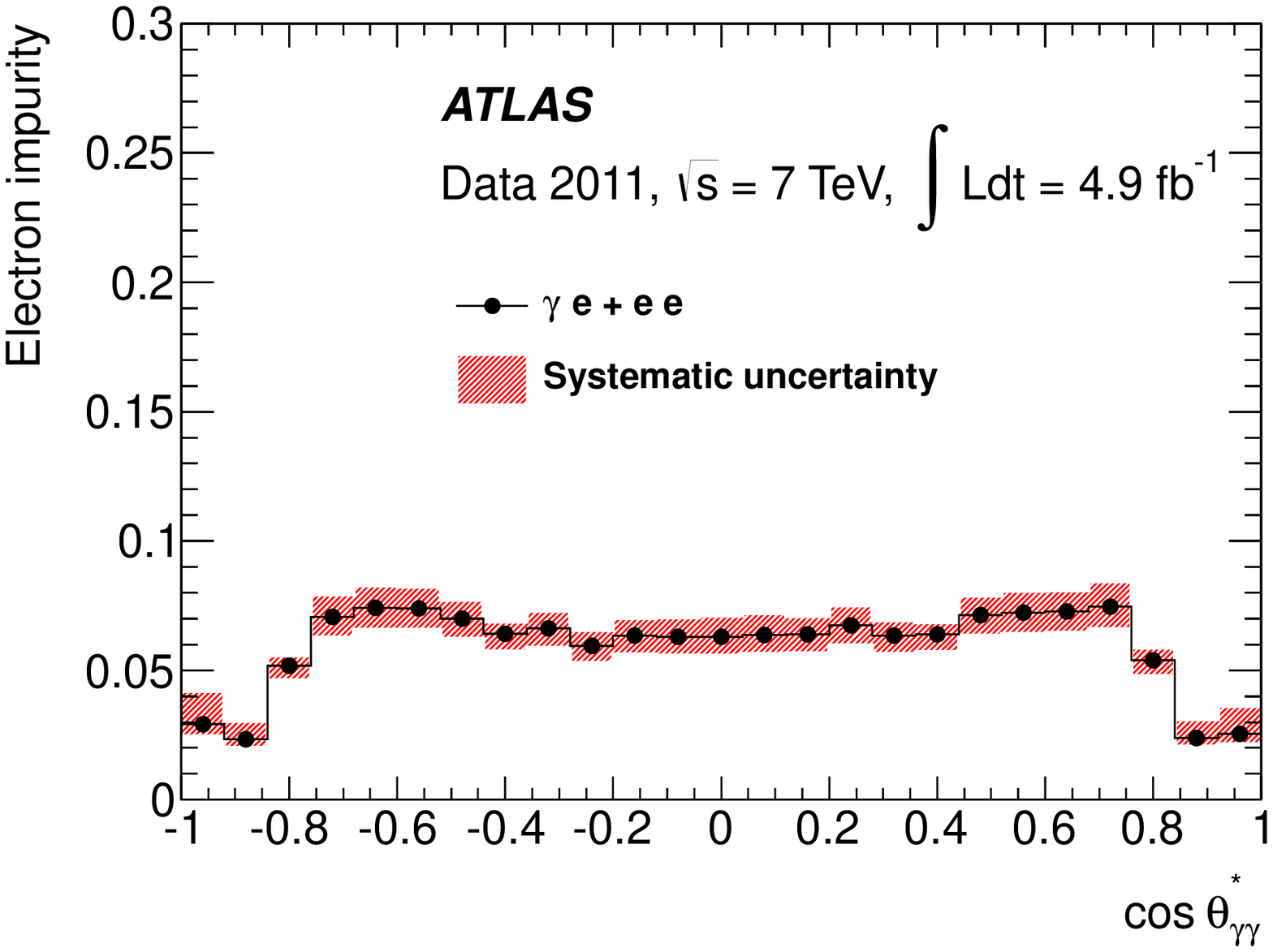}
   \caption{Fraction of electron background (impurity) as a function of \mgg,
     \ptgg, \dphigg, and \cosThetagg.
   }
\label{fig:electron}
 \end{center}
\end{figure}

\section{Cross section measurement}
\label{s:cross_section}
This section describes the extraction of the final cross sections.
The background-subtracted differential spectra are first unfolded to the
generated-particle level, to take into account reconstruction and
selection efficiencies estimated from the simulation,
and then divided by the integrated
luminosity of the data sample 
and the trigger
efficiency relative to the offline selection.

\subsection{Efficiency and unfolding}
\label{ss:efficiency}
The background-subtracted differential distributions
obtained from the data are unfolded to obtain the particle-level
spectra by dividing the signal yield in each bin of the 
\diphoton observable under study by a ``bin-by-bin'' correction, which 
accounts for signal reconstruction and selection efficiencies and
for finite resolution effects.
The bin-by-bin nominal corrections are evaluated from the \sherpa\ 
\diphoton simulated sample as the number of simulated \diphoton events
satisfying the selection criteria (excluding the trigger requirement)
and for which the reconstructed value of the variable $X$ under
consideration is in bin $i$, divided by the number of 
simulated \diphoton events satisfying the nominal acceptance criteria
at generator-level and for which the generated value of $X$ is in the same bin $i$.
The generator-level photon transverse isolation energy is computed from 
the true four-momenta of the generated particles (excluding muons and neutrinos)
inside a cone of radius 0.4 around the photon direction. 
The pileup contribution is removed using an analogous method to the one for 
the experimental isolation variable,
by subtracting the product of the area of the isolation cone and 
the median transverse energy density of the low-transverse-momentum 
truth-particle jets.

Alternative corrections are calculated with the \pythia\ \diphoton sample
or using a simulated \diphoton sample which contains additional 
material upstream of the calorimeter. The variations induced on
the measured cross sections by the alternative corrections are
taken as systematic uncertainties, due to the uncertainty on the
generated kinematic distributions, on the relative fraction of direct and 
fragmentation \diphoton production, and on the amount of material in
the ATLAS detector.
The effect on the total cross section is within
$^{+2}_{-5}$\% for \mgg, $\pm 3$\% for \ptgg, $^{+3}_{-4}$\% for \dphigg\ and
$^{+2}_{-3}$\% for \cosThetagg.

The effect of the uncertainty on the efficiency of the photon
identification criteria is estimated by varying the identification 
efficiency in the simulation by its uncertainty~\cite{ATLAS-CONF-2012-123}. 
The uncertainties on the electromagnetic (photon) energy scale and resolution
are also propagated to the final measurement by varying them within
their uncertainties~\cite{electronperf_2011}.
The effect on the differential cross section is typically $^{+1}_{-2}$\%.
Other uncertainties, related to the dependence on the average number 
of pile-up interactions of the efficiencies of
the photon identification and transverse isolation energy requirements and 
to the observed data--MC shift in the photon transverse isolation 
energy distributions, are found to be negligible.

A closure test has been performed by unfolding the differential spectra of
\diphoton events selected in the \pythia\ signal sample with the bin-by-bin 
coefficients determined using the \sherpa\ sample, and comparing 
the unfolded spectra to the
truth-level spectra in the same \pythia\ sample. Non-closure effects
of at most 2\% have been found and included in the final systematic
uncertainty.

More sophisticated unfolding methods which account for migrations between 
bins, either based on the repeated (iterative) application of Bayes' 
theorem~\cite{iterative-unfolding} or on a least-square minimization 
followed by a regularization of the resulting spectra~\cite{TUnfold}
have also been investigated.
The differences between the unfolded spectra obtained with these
methods and the spectra extracted with the bin-by-bin corrections 
are negligible compared to the other uncertainties and 
therefore the bin-by-bin method was chosen for the final results.

\subsection{Trigger efficiency correction}
The unfolded spectra are then corrected for the event-level trigger
efficiency, defined as the fraction of \diphoton events -- satisfying all
the selection criteria -- that pass the \diphoton trigger used to 
collect the data.
The trigger efficiency is measured in data using a bootstrap technique \cite{Trigger} 
from samples selected with fully efficient unbiased triggers with a lower threshold,
and taking into account kinematic correlations between the photon candidates.
The differences between the measured \diphoton trigger efficiency and
the efficiency estimated with simulated \diphoton samples, or by applying the bootstrap 
technique to single-photon triggers and neglecting correlations between
the two photon candidates, have been assigned as systematic uncertainties.
The total trigger efficiency is then:
\begin{equation}
\label{eq:trig_efficiency}
\varepsilon_{\rm trig} = \left(97.8^{+0.8}_{-1.5} {\rm (stat.)}\pm{0.8} {\rm (syst.)}\right)\%
\end{equation} 
Using \diphoton simulated samples, the trigger efficiency has been estimated
to be constant, within the total uncertainties, 
as a function of the four \diphoton observables under investigation.

\subsection{Results}

The differential cross sections as a function of \mgg, \ptgg, \dphigg, 
and \cosThetagg\ are extracted following the unfolding procedure described 
in section~\ref{ss:efficiency} and using the trigger efficiency quoted in 
eq.~(\ref{eq:trig_efficiency}). The numerical
results are listed in Appendix~\ref{a:xs_tables}.

The integrated cross section is measured by dividing the global
\evtgamgam\ yield (obtained after subtracting the electron
contribution from the two-dimensional fit result in table~\ref{tab:yields})
by the product of the average event selection efficiency (from the
simulation), trigger efficiency and integrated luminosity.
The selection efficiency is defined as the number of
reconstructed simulated \diphoton events satisfying the 
detector-level selection criteria divided by the number of generated
events satisfying the equivalent truth-level criteria, 
thus correcting for reconstructed events with true photons failing 
the acceptance cuts.
It is computed from simulated \diphoton
events, reweighting the spectrum of one of the four \diphoton
variables under study in order to match the differential 
background-subtracted \diphoton spectrum observed in data.
Choosing different variables for the reweighting of the simulated
events leads to slightly different but consistent efficiencies,
with an average value of 49.6\% and an RMS of 0.2\%.
Including systematic uncertainties on the photon reconstruction and 
identification efficiencies, from the same sources described in
section~\ref{ss:efficiency}, the event selection efficiency 
is estimated to be $49.6^{+1.9}_{-1.7}$\%.
The dominant contributions to the efficiency uncertainty are from 
the photon identification efficiency uncertainty ($\pm 1.2\%$), 
the energy scale uncertainty ($^{+1.2}_{-0.5}\%$),
and the choice of the MC generator and the detector simulation ($\pm 0.9\%$).
Negligible uncertainties are found to arise from the energy resolution,
the isolation requirement (evaluated by shifting
the isolation variable by the observed data--MC difference) and from 
the different pile-up dependence of the efficiency in data and MC simulation.
With an integrated luminosity of $(4.9\pm 0.2)$~fb$^{-1}$ at $\sqrt{s} = 7~\tev$,
we obtain an integrated cross section of $44.0^{+3.2}_{-4.2}$~pb, 
where the dominant uncertainties are the event selection
efficiency and the jet subtraction systematic uncertainties. 
As a cross-check, the integrals of the one-dimensional differential
cross sections are also computed. 
They are consistent with the measured integrated
cross section quoted above.

\section{Comparison with theoretical predictions}
\label{s:theory}

The results are compared both to fixed-order NLO and NNLO calculations, 
obtained with parton-level MC generators (\diphox+\gammatwomc\
and \twognnlo), and to the generated-particle-level \diphoton spectra
predicted by leading-order (LO) parton-shower MC generators used in the ATLAS full simulation
(\pythia\ and \sherpa). 
The contribution from the particle recently discovered by 
ATLAS and CMS in the search for the Standard Model Higgs boson
is not included in the predictions: it is expected to 
be around 1\% of the signal in the 120$<\mgg<$130~GeV interval,
and negligible elsewhere.
The contribution from multiple parton interactions is also neglected:
measurements by D0~\cite{Abazov:2011rd} and ATLAS\footnote{article in preparation}
show that events with two jets (in $\gamma$+jets or W+jets) have 
a contamination between 5\% and 10\% from double parton interactions.
In our data sample, the fraction of selected \diphoton\ candidates
with at least two additional jets not overlapping with the photons and
not from pile-up is around 8\%, thus the overall contribution to the signal
from multiple parton interactions is estimated to be lower than 1\%.

The main differences between the four predictions are the following:
\begin{itemize}
\item \twognnlo\ provides a NNLO calculation of the direct part of the \diphoton
production cross section, but neglects completely the contribution from the
fragmentation component, where one or both photons are produced in the 
soft collinear fragmentation of coloured partons.
\item \diphox\ provides a NLO calculation of both the direct and the
fragmentation parts of the \diphoton production cross section. It also
includes the contribution from the box diagram ($gg \to \gamma\gamma$), 
which is in principle a term of the NNLO expansion in the strong
coupling constant $\alpha_s$, but -- due to the large gluon luminosity
at the LHC~\cite{gluon_luminosity} -- gives a contribution comparable
to that of the LO terms. 
For these reasons, higher-order contributions to the box
diagrams, technically at NNNLO but of size similar to that of 
NLO terms, are also included in our calculation by using \gammatwomc.
\item \pythia\ provides LO matrix elements for \diphoton production and
models the higher-order terms through \gamjet\ and di-jet production in
combination with initial-state and/or final-state radiation.
It also features parton showering and an underlying event model;
\item \sherpa\ has features similar to those of \pythia, and in addition
includes the \diphoton higher-order real-emission matrix elements.
For this study, up to two additional QCD partons are generated.
\end{itemize}

The nominal factorization ($\mu_F$), renormalization ($\mu_R$), and --
in the case of \diphox\ and \gammatwomc\ -- fragmentation ($\mu_f$) scales
are set in all cases to the \diphoton invariant mass, $\mgg$. 
Different PDF sets are used by each program: CT10 NLO~\cite{ct10} for
\diphox\ and \gammatwomc, MSTW2008 NNLO~\cite{mstw} for \twognnlo, 
CTEQ6L1 for \sherpa\ and MRST2007 LO* for \pythia.
The theoretical uncertainty error bands for \pythia\ and \sherpa\ include
only statistical uncertainties.
The theory uncertainty error bands for the NLO and NNLO predictions
include in addition PDF and scale uncertainties. PDF uncertainties are estimated
by varying each of the eigenvalues of the PDFs by $\pm 1\sigma$ and
summing in quadrature separately positive and negative variations of
the cross section. 
For \diphox\ and \gammatwomc, scale uncertainties are 
evaluated by varying each scale to $\mgg/2$ and $2\mgg$, and the envelope
of all variations is taken as a systematic error; the final
uncertainty is dominated by the configurations in which the scales are
varied incoherently. For \twognnlo, the scale uncertainty is evaluated 
by considering the variation of the
predicted cross sections in the two cases 
$\mu_R = \mgg/2$, $\mu_F = 2\mgg$ and $\mu_R = 2\mgg$, $\mu_F = \mgg/2$.

Fixed-order predictions calculated at parton level 
do not include underlying event, pile-up or hadronization effects. 
While the ambient-energy density corrections to the photon isolation
are expected to remove most of these effects from the photon isolation energy,
it is not guaranteed that they correct the experimental isolation back
to exactly the parton-level isolation computed from the
elementary-process partons. 
To estimate these residual effects, \pythia\ and \sherpa\ \diphoton
samples are used to evaluate the ratio of generator-level cross sections
with and without hadronization and the underlying event, and subsequently, 
the parton-level cross sections are multiplied bin-by-bin by this ratio.
The central value of the envelope of the \pythia\ and \sherpa\
distributions is taken as the nominal correction and half
of the difference between \pythia\ and \sherpa\, as the systematic
uncertainty. The typical correction factor is around 0.95.

Both \pythia\ and \sherpa\ are expected to underestimate the
total cross section, because of the missing NLO (and higher-order) 
contributions.
At low \ptgg\ and for \dphigg\ near $\pi$
where multiple soft gluon emission is important, \pythia\
and \sherpa\ are expected to better describe the
shape of the differential distributions, thanks to the effective 
all-order resummation of the leading logs performed by the parton
shower. On the other hand, in the same regions fixed-order
calculations are expected to exhibit infrared divergences. 
Finally, \twognnlo\ is expected to underestimate the data 
in regions populated by the contribution from fragmentation
(low \dphigg\ and \mgg, and \cosThetagg$\approx$1).

The total cross section estimated by \pythia\ and \sherpa\ with the
ATLAS simulation settings is 36 pb, and underestimates the measured 
cross section by 20\%.
The \diphox+\gammatwomc\ total cross section is
$39^{+7}_{-6}$ pb and the \twognnlo\ total cross section is $44^{+6}_{-5}$ pb,
where the uncertainty is dominated by the choice of the nominal scales.

The comparisons between the experimental cross sections and the predictions
by \pythia\ and \sherpa\ are shown in
figure~\ref{f:data_theory_xsection_comparison_LO}.
In order to compare the shapes of the MC differential distributions
to the data, their cross sections are rescaled by a factor 1.2
to match the total cross section measured in data.
\pythia\ misses higher order contributions, as clearly seen for low
values of \dphigg, but this is compensated by the parton shower
for \dphigg\ near $\pi$ and at low \ptgg. It is worth noting that the
shoulder expected (and observed) in the \ptgg\ cross section around
the sum of the $\ET$ thresholds of the two
photons~\cite{guillet_shoulder} is almost absent in \pythia,
while \sherpa\ correctly reproduces the data in this region.
This is interpreted as being due to the additional NLO
contributions in \sherpa\ combined with differences in the parton
showers. Overall, \sherpa\ reproduces  the data rather well,
except at large \mgg\ and large $|\cosThetagg|$.

\begin{figure}[!htbp]
 \begin{center}
   \includegraphics[width=0.48\columnwidth]{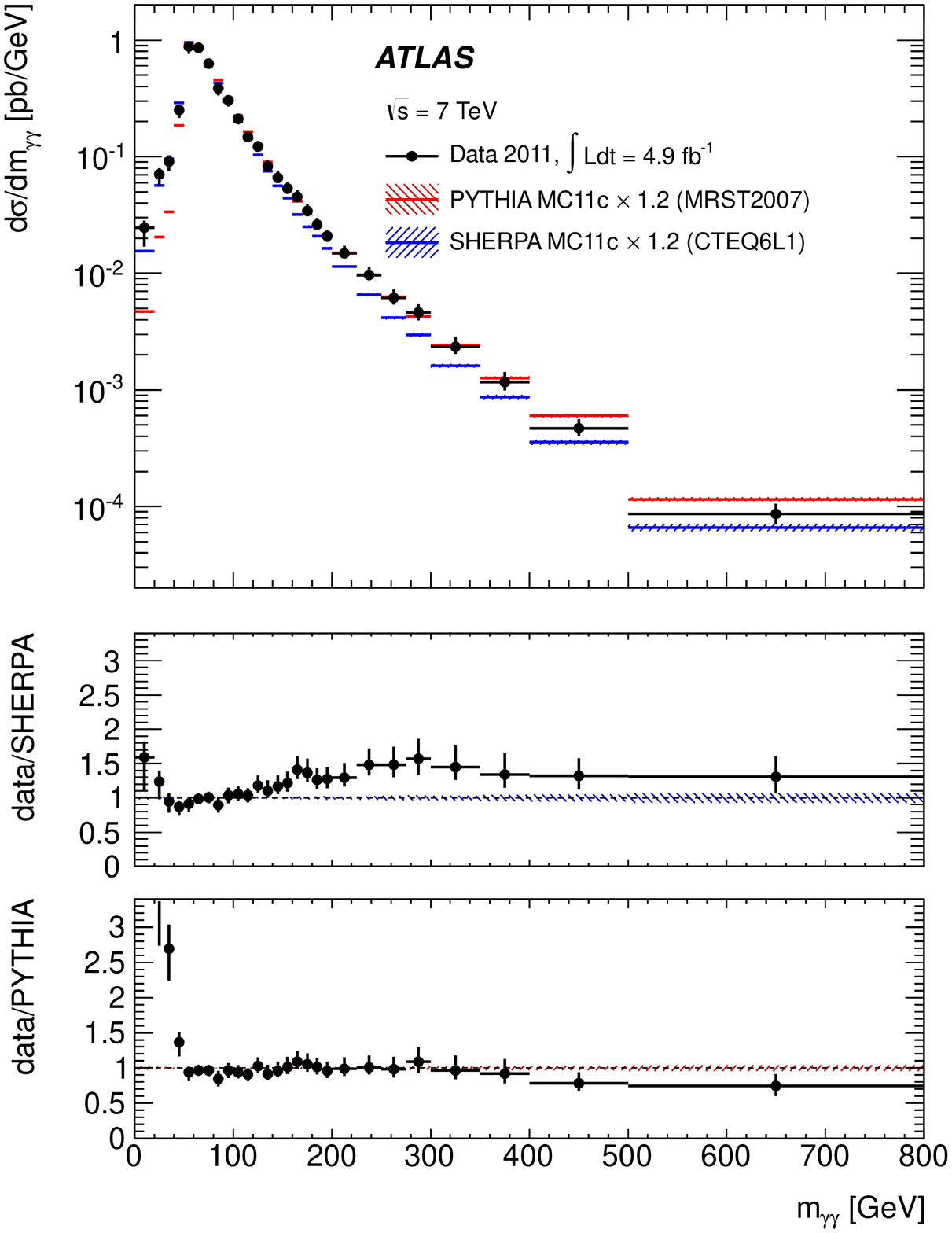} 
   \includegraphics[width=0.48\columnwidth]{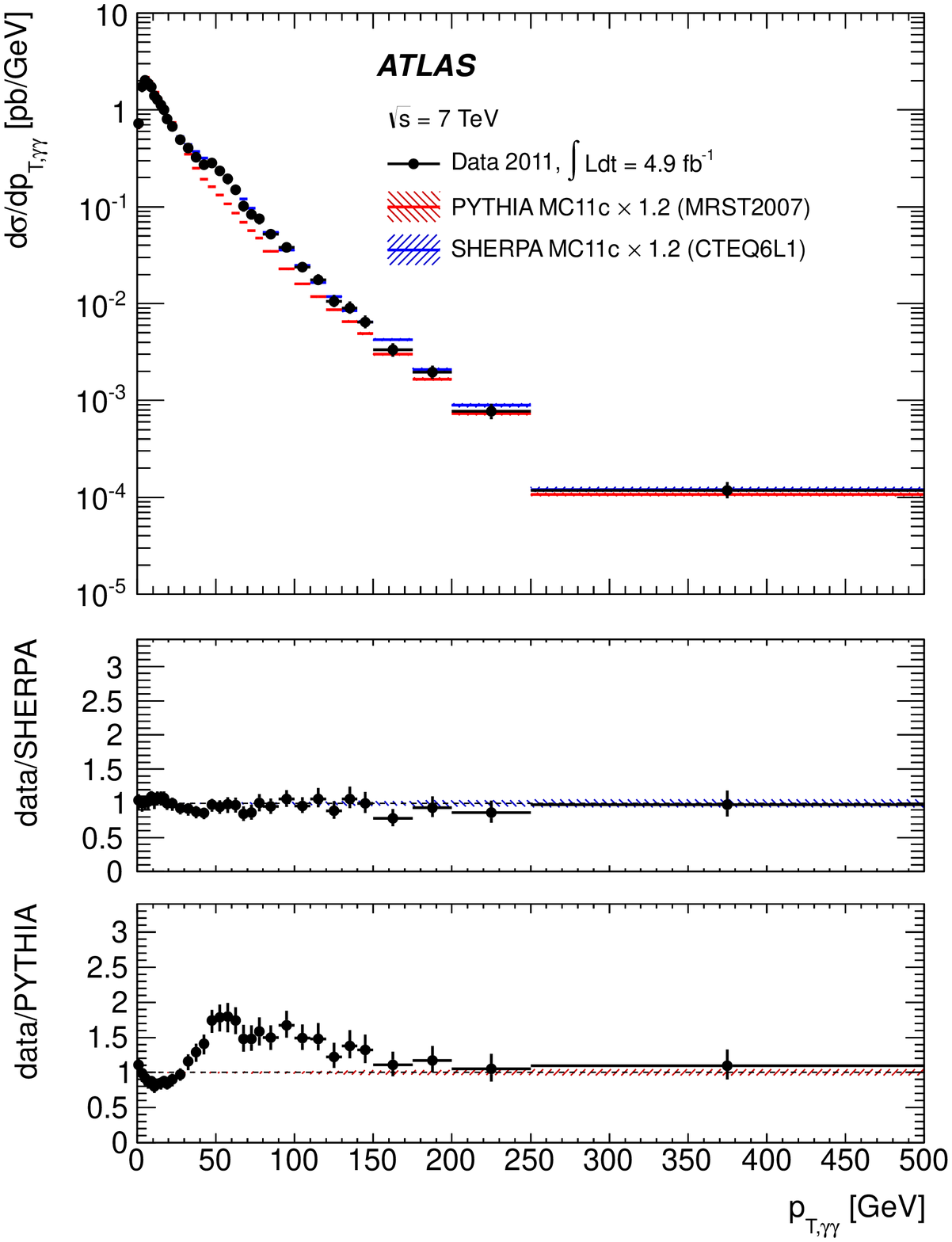}
   \includegraphics[width=0.48\columnwidth]{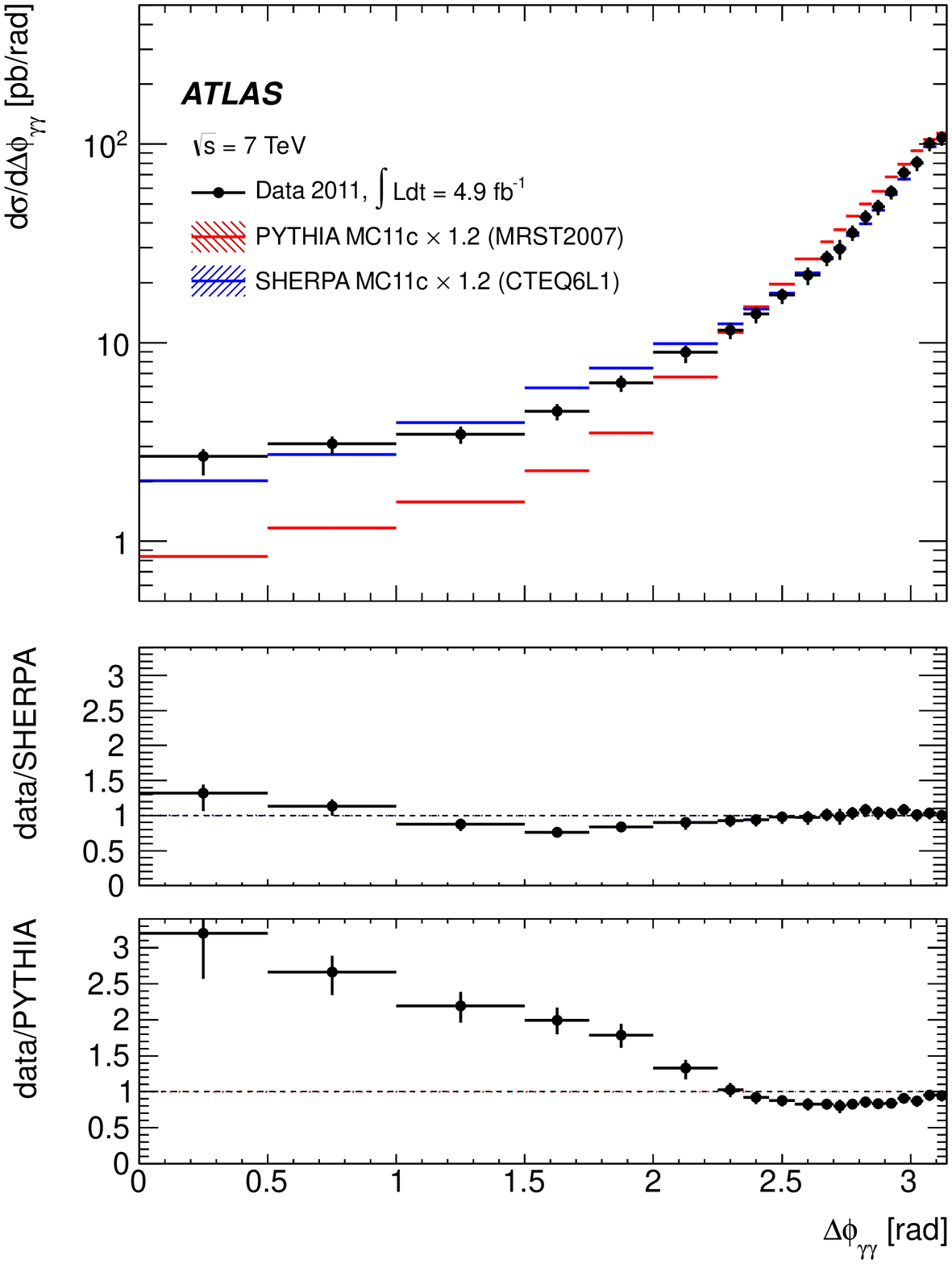}
   \includegraphics[width=0.48\columnwidth]{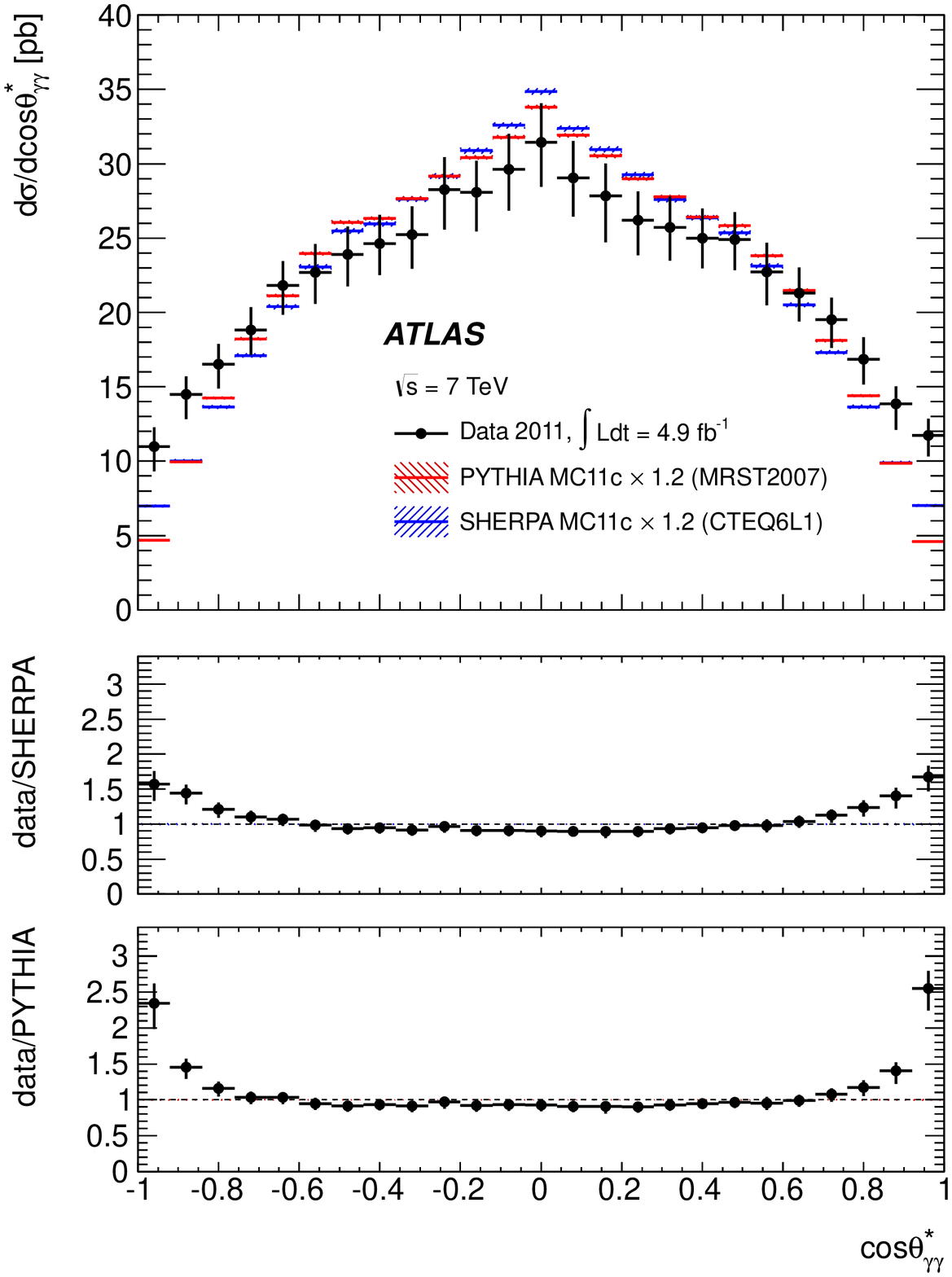}
   \caption{Comparison between the experimental cross sections 
    and the predictions obtained with parton-shower LO simulations:
    \mgg~(top left), \ptgg~(top right),
    \dphigg~(bottom left), \cosThetagg (bottom right).
    The LO cross sections have been scaled to the total data cross section,
    by a factor 1.2.
    Black dots correspond to data with error bars for their total uncertainties,
    which are dominated by the systematic component.
    The simulated cross sections include only statistical uncertainties.
}
  \label{f:data_theory_xsection_comparison_LO}
 \end{center}
\end{figure}

The comparisons between the data cross sections and the 
predictions by \twognnlo\ and \diphox+\gammatwomc\ are shown in
figure~\ref{f:data_theory_xsection_comparison_NLO}. In the
$\dphigg\simeq\pi$, low \ptgg\  region, \diphox+\gammatwomc\ fails to
match the data. This is expected because initial-state soft
gluon radiation is divergent at NLO, without soft gluon
resummation. Everywhere else \diphox+\gammatwomc\ is missing NNLO
contributions and clearly underestimates the data. 

With higher order calculations included, \twognnlo\ is very close to
the data within the uncertainties. However, the excess at
$\dphigg\simeq\pi$ and low \ptgg\ is still present, as expected for 
a fixed-order calculation.
Since the fragmentation component is not calculated in
\twognnlo,  the data is slightly underestimated by \twognnlo\
in the regions where this component is larger: 
at low \dphigg, low mass, intermediate
\ptgg\ (between 20 GeV and 150 GeV) and large $|\cosThetagg|$.

\begin{figure}[!htbp]
 \begin{center}
   \includegraphics[width=0.48\columnwidth]{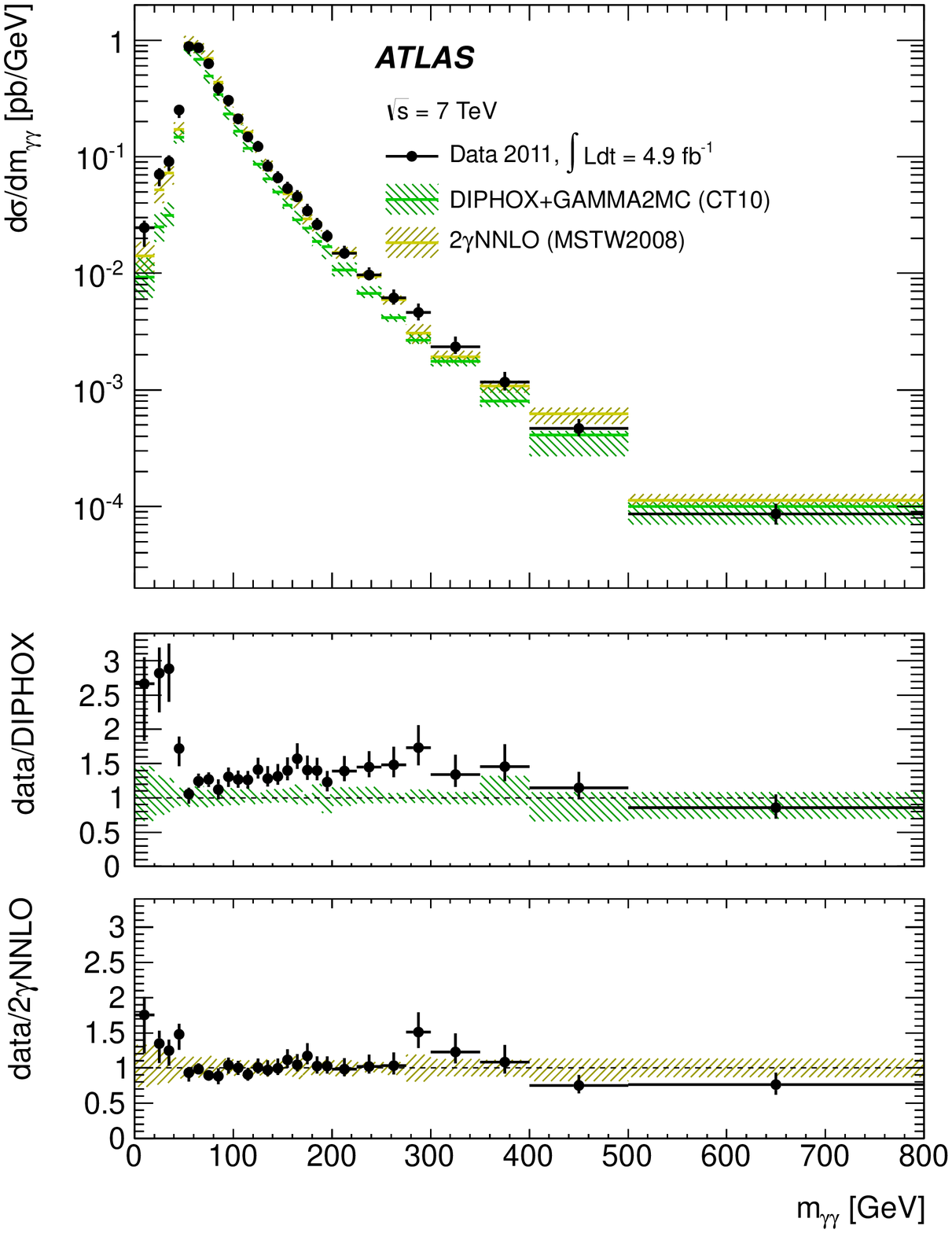} 
   \includegraphics[width=0.48\columnwidth]{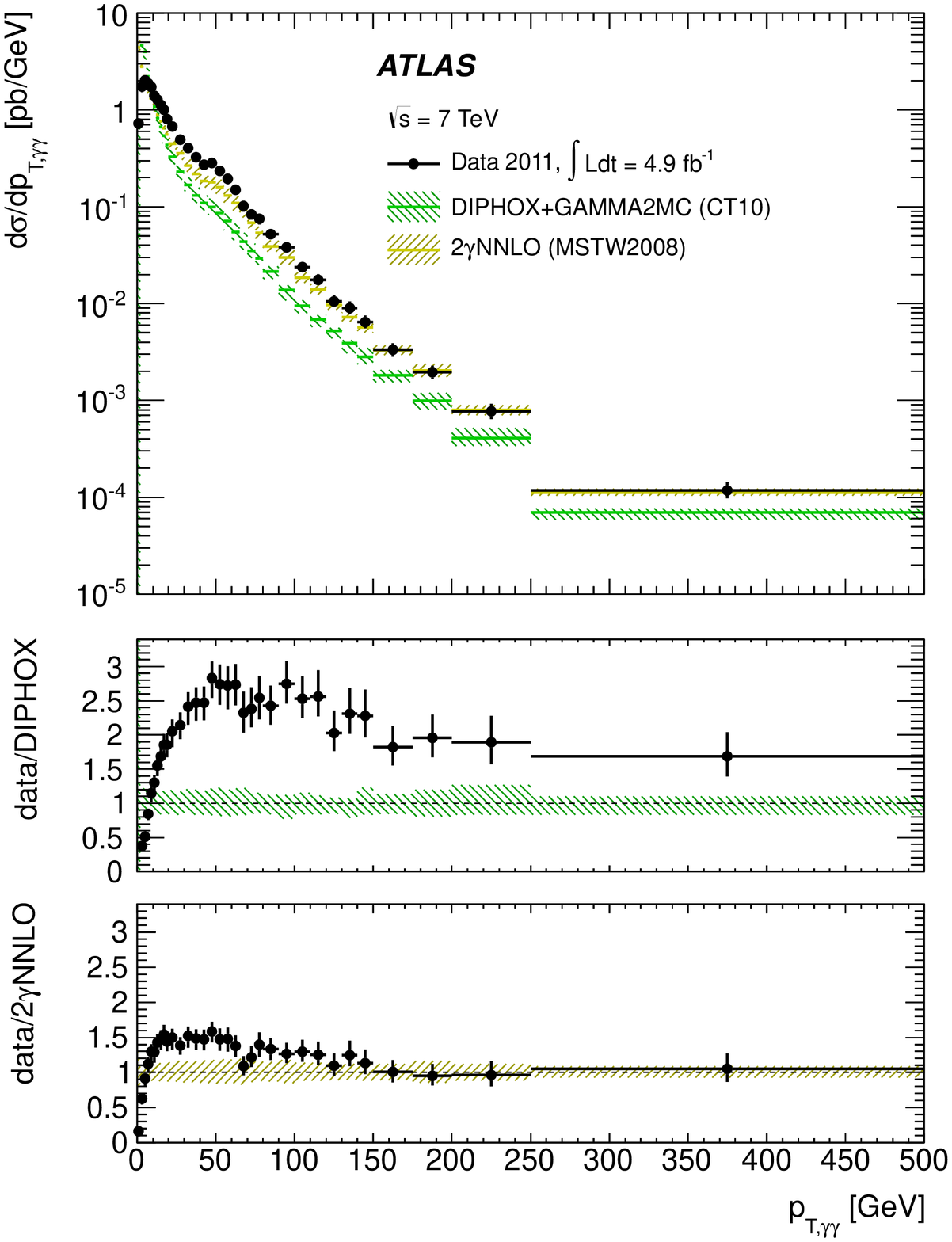}
   \includegraphics[width=0.48\columnwidth]{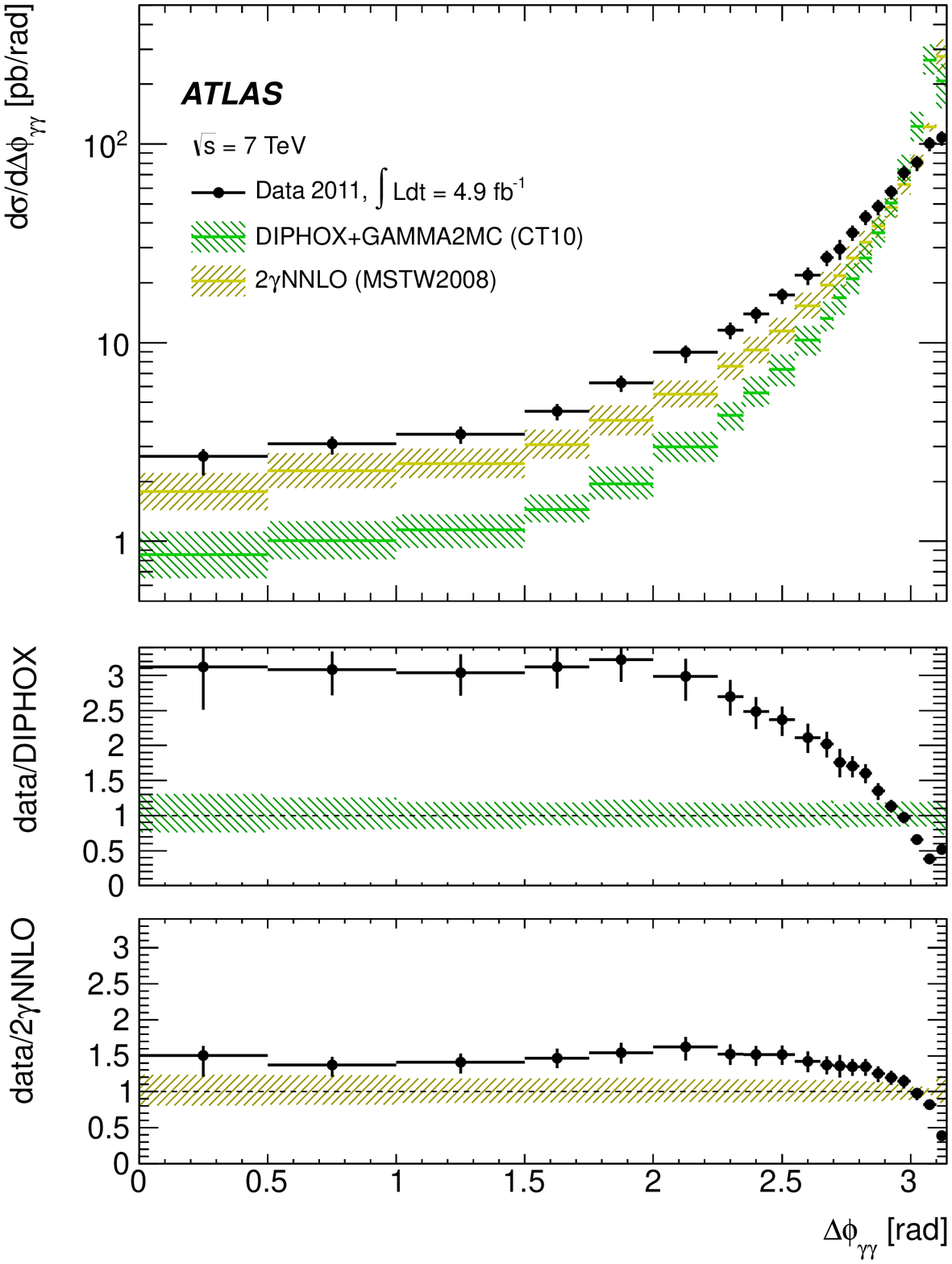}
   \includegraphics[width=0.48\columnwidth]{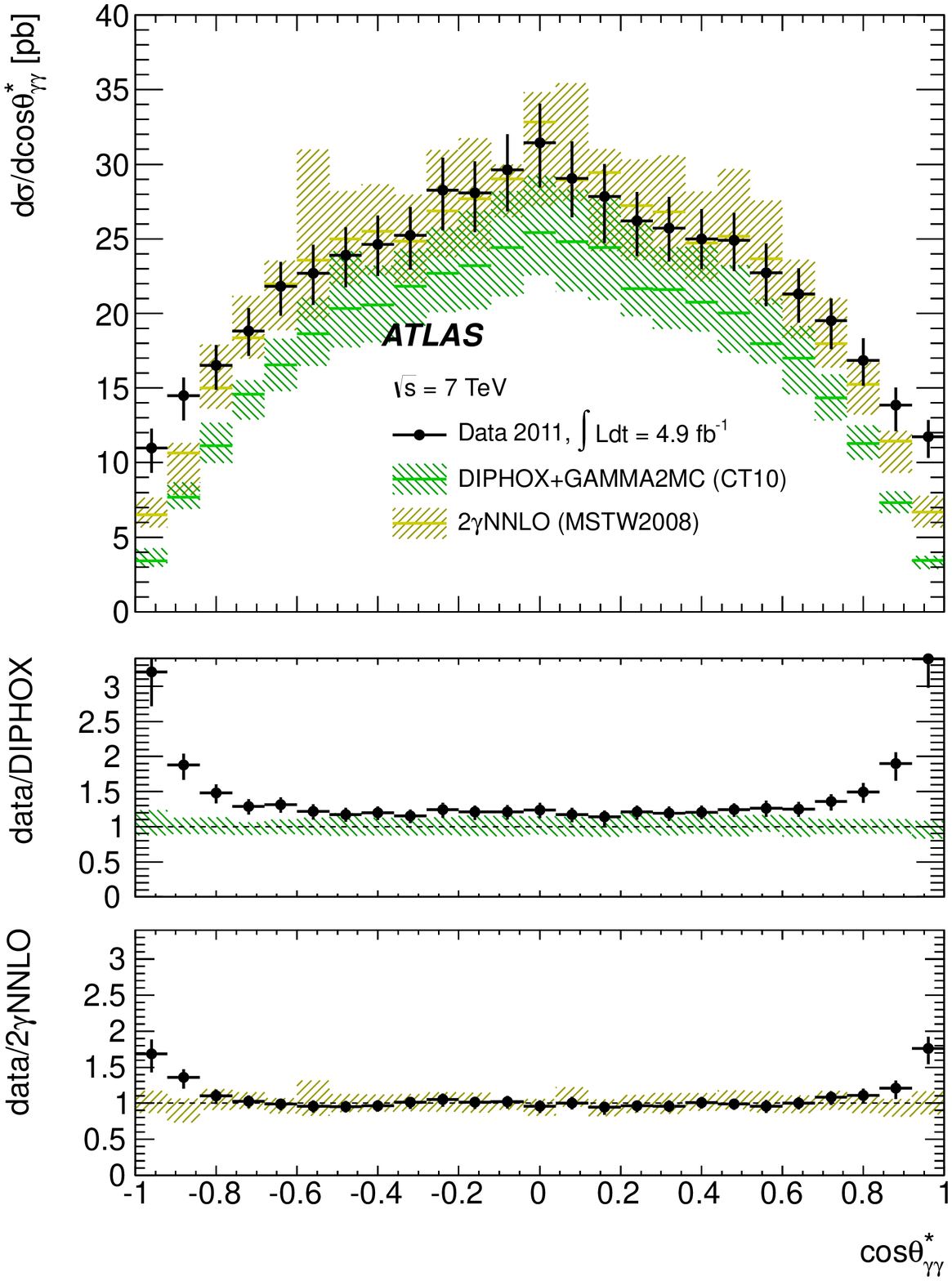}
   \caption{Comparison between the experimental cross sections 
    and the predictions obtained with \diphox+\gammatwomc\ (NLO) and
    \twognnlo\ (NNLO):
    \mgg~(top left), \ptgg~(top right),
    \dphigg~(bottom left), \cosThetagg (bottom right).
    Black dots correspond to data with with error bars for their total uncertainties,
    which are dominated by the systematic component.
    The theoretical uncertainties include contributions from the 
    limited size of the simulated sample, from the scale
    choice and from uncertainties on the parton distribution functions and 
    on the hadronization and underlying event corrections.
   }
  \label{f:data_theory_xsection_comparison_NLO}
 \end{center}
\end{figure}

\section{Conclusion}
\label{s:conclusion}
A measurement of the production cross section of 
isolated-photon pairs in $pp$ collisions at
a centre-of-mass energy $\sqrt{s} = 7\TeV$ is presented.
The measurement uses an integrated luminosity of 4.9~\ifb\
collected by the ATLAS detector at the LHC in 2011.
The two photons are required to be isolated in the calorimeters,
to be in the acceptance of
the electromagnetic calorimeter ($|\eta|<2.37$ with the
exclusion of the barrel-endcap transition region $1.37<|\eta|<1.52$)
and to have an angular separation $\Delta R>0.4$ in the
$\eta,\phi$ plane. 
Both photons have transverse energies $\ET>22$~GeV, and at least
one of them has ${\ET>25}$~GeV.

The total cross section within the acceptance is $44.0^{+3.2}_{-4.2}$~pb.
It is underestimated by \sherpa\ and \pythia,
which both predict a value of 36 pb with the current ATLAS simulation tune.
The central value of the cross section predicted by \diphox+\gammatwomc,
39 pb, is lower than the data but it is consistent with data within
the theoretical ($^{+7}_{-6}$ pb) and experimental errors.
The NNLO calculation of \twognnlo\ ($\sigma_{\rm NNLO} = 44^{+6}_{-5}$~pb)
is in excellent agreement with the data.

The differential cross sections, as a function of the \diphoton
invariant mass, transverse momentum, azimuthal separation and
of the cosine of the polar angle of the photon with largest transverse energy in the
Collins--Soper \diphoton rest frame, are also measured.
Rather good agreement is found with Monte Carlo generators, after
rescaling the \pythia\ and \sherpa\ distributions by a factor 1.2
in order to match the integrated cross section measured in data
and fixed-order calculations, in the regions of phase space studied. 
All generators tend to underestimate the data at large 
$|\cos\theta^*_{\gamma\gamma}|$.
\sherpa\ performs rather well for most differential spectra, except 
for high \mgg.
\pythia\ is missing higher order contributions, but this is compensated 
by the parton shower for \dphigg\ near $\pi$ and at low \ptgg. 
In these same regions the fixed-order calculations do not reproduce 
the data, due to the known infrared divergences from initial-state 
soft gluon radiation.
Everywhere else \diphox+\gammatwomc\ is missing NNLO
contributions and clearly underestimates the data. On the other hand, with
inclusion of NNLO terms, \twognnlo\ is able to match the data very closely 
within the uncertainties, except in limited regions where the fragmentation 
component -- neglected in the \twognnlo\ calculation -- is still significant
after the photon isolation requirement.

\acknowledgments

We thank the 2$\gamma$NNLO authors for providing theoretical
predictions of the NNLO di-photon cross section based on their
code. We thank Leandro Cieri, Stefano Catani, Daniel De Florian, 
Michel Fontannaz, Jean-Philippe Guillet, Eric Pilon and Carl Schmidt,
for fruitful discussions on the theoretical calculations and their
uncertainties.  

We thank CERN for the very successful operation of the LHC, as well as the
support staff from our institutions without whom ATLAS could not be
operated efficiently.

We acknowledge the support of ANPCyT, Argentina; YerPhI, Armenia; ARC,
Australia; BMWF and FWF, Austria; ANAS, Azerbaijan; SSTC, Belarus; CNPq and FAPESP,
Brazil; NSERC, NRC and CFI, Canada; CERN; CONICYT, Chile; CAS, MOST and NSFC,
China; COLCIENCIAS, Colombia; MSMT CR, MPO CR and VSC CR, Czech Republic;
DNRF, DNSRC and Lundbeck Foundation, Denmark; EPLANET, ERC and NSRF, European Union;
IN2P3-CNRS, CEA-DSM/IRFU, France; GNSF, Georgia; BMBF, DFG, HGF, MPG and AvH
Foundation, Germany; GSRT and NSRF, Greece; ISF, MINERVA, GIF, DIP and Benoziyo Center,
Israel; INFN, Italy; MEXT and JSPS, Japan; CNRST, Morocco; FOM and NWO,
Netherlands; BRF and RCN, Norway; MNiSW, Poland; GRICES and FCT, Portugal; MERYS
(MECTS), Romania; MES of Russia and ROSATOM, Russian Federation; JINR; MSTD,
Serbia; MSSR, Slovakia; ARRS and MVZT, Slovenia; DST/NRF, South Africa;
MICINN, Spain; SRC and Wallenberg Foundation, Sweden; SER, SNSF and Cantons of
Bern and Geneva, Switzerland; NSC, Taiwan; TAEK, Turkey; STFC, the Royal
Society and Leverhulme Trust, United Kingdom; DOE and NSF, United States of
America.

The crucial computing support from all WLCG partners is acknowledged
gratefully, in particular from CERN and the ATLAS Tier-1 facilities at
TRIUMF (Canada), NDGF (Denmark, Norway, Sweden), CC-IN2P3 (France),
KIT/GridKA (Germany), INFN-CNAF (Italy), NL-T1 (Netherlands), PIC (Spain),
ASGC (Taiwan), RAL (UK) and BNL (USA) and in the Tier-2 facilities
worldwide.


\bibliographystyle{JHEP}
\providecommand{\href}[2]{#2}\begingroup\raggedright\endgroup

\clearpage
\appendix

\section{Experimental differential cross section}
\label{a:xs_tables}

The numerical values of the differential cross sections displayed in figures~\ref{f:data_theory_xsection_comparison_LO}
and \ref{f:data_theory_xsection_comparison_NLO} are quoted
in tables~\ref{tab:cross_section_mgg}-\ref{tab:cross_section_costh}. For each bin of
the \mgg, \ptgg, \dphigg, and \cosThetagg~ variables, the cross section is given together with its
statistical, systematic and total uncertainties. All values are divided by the bin width.

\begin{table}[!hb]
  \begin{center}
\begin{tabular}{lrrrrrrr}
\hline \hline
 \mgg  & d$\sigma$/d\mgg & \multicolumn{2}{c}{Statistical error} & \multicolumn{2}{c}{Systematic errors} & \multicolumn{2}{c}{Total error} \\
 $[$GeV$]$ & [pb/GeV] & high  & low & high & low & high & low \\
\hline
 $[0,20)$ & $0.0247$ & $+0.0015$ & $-0.0015$ & $+0.0032$ & $-0.0076$ & $+0.0036$ & $-0.0077$ \\
 $[20,30)$ & $0.0704$ & $+0.0032$ & $-0.0032$ & $+0.0087$ & $-0.0140$ & $+0.0093$ & $-0.0144$ \\
 $[30,40)$ & $0.091$ & $+0.004$ & $-0.004$ & $+0.011$ & $-0.015$ & $+0.012$ & $-0.015$ \\
 $[40,50)$ & $0.252$ & $+0.006$ & $-0.006$ & $+0.025$ & $-0.037$ & $+0.026$ & $-0.037$ \\
 $[50,60)$ & $0.880$ & $+0.012$ & $-0.012$ & $+0.073$ & $-0.120$ & $+0.074$ & $-0.120$ \\
 $[60,70)$ & $0.857$ & $+0.010$ & $-0.010$ & $+0.071$ & $-0.068$ & $+0.071$ & $-0.068$ \\
 $[70,80)$ & $0.626$ & $+0.008$ & $-0.008$ & $+0.051$ & $-0.052$ & $+0.051$ & $-0.053$ \\
 $[80,90)$ & $0.384$ & $+0.008$ & $-0.008$ & $+0.049$ & $-0.048$ & $+0.050$ & $-0.049$ \\
 $[90,100)$ & $0.305$ & $+0.006$ & $-0.006$ & $+0.031$ & $-0.034$ & $+0.032$ & $-0.035$ \\
 $[100,110)$ & $0.212$ & $+0.004$ & $-0.004$ & $+0.021$ & $-0.021$ & $+0.021$ & $-0.021$ \\
 $[110,120)$ & $0.148$ & $+0.004$ & $-0.004$ & $+0.015$ & $-0.014$ & $+0.015$ & $-0.015$ \\
 $[120,130)$ & $0.122$ & $+0.003$ & $-0.003$ & $+0.015$ & $-0.010$ & $+0.015$ & $-0.011$ \\
 $[130,140)$ & $0.0829$ & $+0.0025$ & $-0.0025$ & $+0.0112$ & $-0.0073$ & $+0.0115$ & $-0.0077$ \\
 $[140,150)$ & $0.0656$ & $+0.0022$ & $-0.0022$ & $+0.0088$ & $-0.0057$ & $+0.0091$ & $-0.0061$ \\
 $[150,160)$ & $0.0535$ & $+0.0019$ & $-0.0019$ & $+0.0072$ & $-0.0051$ & $+0.0075$ & $-0.0054$ \\
 $[160,170)$ & $0.0451$ & $+0.0017$ & $-0.0017$ & $+0.0063$ & $-0.0038$ & $+0.0065$ & $-0.0042$ \\
 $[170,180)$ & $0.0343$ & $+0.0015$ & $-0.0015$ & $+0.0050$ & $-0.0031$ & $+0.0052$ & $-0.0035$ \\
 $[180,190)$ & $0.0262$ & $+0.0013$ & $-0.0013$ & $+0.0032$ & $-0.0024$ & $+0.0035$ & $-0.0027$ \\
 $[190,200)$ & $0.0209$ & $+0.0011$ & $-0.0011$ & $+0.0025$ & $-0.0019$ & $+0.0028$ & $-0.0022$ \\
 $[200,225)$ & $0.0149$ & $+0.0006$ & $-0.0006$ & $+0.0023$ & $-0.0014$ & $+0.0024$ & $-0.0015$ \\
 $[225,250)$ & $0.00970$ & $+0.00049$ & $-0.00049$ & $+0.00150$ & $-0.00086$ & $+0.00158$ & $-0.00099$ \\
 $[250,275)$ & $0.00616$ & $+0.00039$ & $-0.00039$ & $+0.00104$ & $-0.00064$ & $+0.00111$ & $-0.00075$ \\
 $[275,300)$ & $0.00464$ & $+0.00036$ & $-0.00036$ & $+0.00080$ & $-0.00059$ & $+0.00087$ & $-0.00069$ \\
 $[300,350)$ & $0.00235$ & $+0.00017$ & $-0.00017$ & $+0.00048$ & $-0.00026$ & $+0.00051$ & $-0.00031$ \\
 $[350,400)$ & $0.00116$ & $+0.00011$ & $-0.00011$ & $+0.00024$ & $-0.00013$ & $+0.00026$ & $-0.00017$ \\
 $[400,500)$ & $4.69$e$-04$ & $+5.0$e$-05$ & $-5.0$e$-05$ & $+7.9$e$-05$ & $-4.7$e$-05$ & $+9.3$e$-05$ & $-6.9$e$-05$ \\
 $[500,800)$ & $8.6$e$-05$ & $+1.3$e$-05$ & $-1.3$e$-05$ & $+1.5$e$-05$ & $-1.0$e$-05$ & $+1.9$e$-05$ & $-1.6$e$-05$ \\
\hline \hline
\end{tabular}
    \caption{
      Experimental cross-section values per bin in pb/GeV for \mgg.
      The listed total errors are the quadratic sum of statistical and systematic uncertainties.
    }
    \label{tab:cross_section_mgg}
  \end{center}
\end{table}

\begin{table}[!hb]
  \begin{center}
\begin{tabular}{lrrrrrrr}
\hline \hline
 \ptgg  & d$\sigma$/d\ptgg & \multicolumn{2}{c}{Statistical error} & \multicolumn{2}{c}{Systematic errors} & \multicolumn{2}{c}{Total error} \\
 $[$GeV$]$ & [pb/GeV] & high  & low & high & low & high & low \\
\hline
 $[0,2)$ & $0.727$ & $+0.022$ & $-0.022$ & $+0.057$ & $-0.092$ & $+0.061$ & $-0.094$ \\
 $[2,4)$ & $1.75$ & $+0.04$ & $-0.04$ & $+0.13$ & $-0.23$ & $+0.13$ & $-0.23$ \\
 $[4,6)$ & $2.03$ & $+0.04$ & $-0.04$ & $+0.15$ & $-0.23$ & $+0.15$ & $-0.23$ \\
 $[6,8)$ & $1.88$ & $+0.04$ & $-0.04$ & $+0.15$ & $-0.21$ & $+0.16$ & $-0.21$ \\
 $[8,10)$ & $1.72$ & $+0.03$ & $-0.03$ & $+0.13$ & $-0.19$ & $+0.14$ & $-0.19$ \\
 $[10,12)$ & $1.40$ & $+0.03$ & $-0.03$ & $+0.12$ & $-0.16$ & $+0.12$ & $-0.16$ \\
 $[12,14)$ & $1.28$ & $+0.03$ & $-0.03$ & $+0.10$ & $-0.13$ & $+0.11$ & $-0.13$ \\
 $[14,16)$ & $1.122$ & $+0.026$ & $-0.026$ & $+0.093$ & $-0.114$ & $+0.097$ & $-0.117$ \\
 $[16,18)$ & $0.999$ & $+0.024$ & $-0.024$ & $+0.086$ & $-0.090$ & $+0.090$ & $-0.093$ \\
 $[18,20)$ & $0.810$ & $+0.021$ & $-0.021$ & $+0.072$ & $-0.076$ & $+0.075$ & $-0.079$ \\
 $[20,25)$ & $0.674$ & $+0.012$ & $-0.012$ & $+0.056$ & $-0.074$ & $+0.058$ & $-0.075$ \\
 $[25,30)$ & $0.492$ & $+0.011$ & $-0.011$ & $+0.041$ & $-0.045$ & $+0.043$ & $-0.047$ \\
 $[30,35)$ & $0.405$ & $+0.009$ & $-0.009$ & $+0.034$ & $-0.043$ & $+0.035$ & $-0.044$ \\
 $[35,40)$ & $0.325$ & $+0.009$ & $-0.009$ & $+0.028$ & $-0.034$ & $+0.030$ & $-0.035$ \\
 $[40,45)$ & $0.272$ & $+0.008$ & $-0.008$ & $+0.024$ & $-0.027$ & $+0.026$ & $-0.028$ \\
 $[45,50)$ & $0.282$ & $+0.008$ & $-0.008$ & $+0.023$ & $-0.027$ & $+0.024$ & $-0.028$ \\
 $[50,55)$ & $0.235$ & $+0.007$ & $-0.007$ & $+0.023$ & $-0.025$ & $+0.025$ & $-0.026$ \\
 $[55,60)$ & $0.194$ & $+0.006$ & $-0.006$ & $+0.019$ & $-0.024$ & $+0.021$ & $-0.024$ \\
 $[60,65)$ & $0.150$ & $+0.006$ & $-0.006$ & $+0.015$ & $-0.016$ & $+0.016$ & $-0.017$ \\
 $[65,70)$ & $0.102$ & $+0.005$ & $-0.005$ & $+0.013$ & $-0.012$ & $+0.014$ & $-0.013$ \\
 $[70,75)$ & $0.0836$ & $+0.0041$ & $-0.0041$ & $+0.0103$ & $-0.0087$ & $+0.0111$ & $-0.0096$ \\
 $[75,80)$ & $0.0748$ & $+0.0036$ & $-0.0036$ & $+0.0087$ & $-0.0086$ & $+0.0094$ & $-0.0093$ \\
 $[80,90)$ & $0.0521$ & $+0.0021$ & $-0.0021$ & $+0.0059$ & $-0.0056$ & $+0.0063$ & $-0.0059$ \\
 $[90,100)$ & $0.0381$ & $+0.0017$ & $-0.0017$ & $+0.0043$ & $-0.0036$ & $+0.0047$ & $-0.0040$ \\
 $[100,110)$ & $0.0239$ & $+0.0013$ & $-0.0013$ & $+0.0028$ & $-0.0023$ & $+0.0031$ & $-0.0026$ \\
 $[110,120)$ & $0.0175$ & $+0.0011$ & $-0.0011$ & $+0.0024$ & $-0.0016$ & $+0.0027$ & $-0.0019$ \\
 $[120,130)$ & $0.0106$ & $+0.0009$ & $-0.0009$ & $+0.0015$ & $-0.0011$ & $+0.0017$ & $-0.0014$ \\
 $[130,140)$ & $0.0090$ & $+0.0008$ & $-0.0008$ & $+0.0012$ & $-0.0008$ & $+0.0015$ & $-0.0012$ \\
 $[140,150)$ & $0.00646$ & $+0.00064$ & $-0.00064$ & $+0.00089$ & $-0.00063$ & $+0.00110$ & $-0.00090$ \\
 $[150,175)$ & $0.00333$ & $+0.00031$ & $-0.00031$ & $+0.00047$ & $-0.00039$ & $+0.00056$ & $-0.00049$ \\
 $[175,200)$ & $0.00195$ & $+0.00023$ & $-0.00023$ & $+0.00025$ & $-0.00017$ & $+0.00034$ & $-0.00028$ \\
 $[200,250)$ & $0.00077$ & $+0.00010$ & $-0.00010$ & $+0.00012$ & $-0.00008$ & $+0.00016$ & $-0.00013$ \\
 $[250,500)$ & $1.18$e$-04$ & $+1.7$e$-05$ & $-1.7$e$-05$ & $+1.8$e$-05$ & $-1.2$e$-05$ & $+2.5$e$-05$ & $-2.1$e$-05$ \\
\hline \hline
\end{tabular}
    \caption{
      Experimental cross-section values per bin in pb/GeV for \ptgg.
      The listed total errors are the quadratic sum of statistical and systematic uncertainties.
    }
    \label{tab:cross_section_ptgg}
  \end{center}
\end{table}

\begin{table}[!hb]
  \begin{center}
\begin{tabular}{lrrrrrrr}
\hline \hline
 \dphigg  & d$\sigma$/d\dphigg & \multicolumn{2}{c}{Statistical error} & \multicolumn{2}{c}{Systematic errors} & \multicolumn{2}{c}{Total error} \\
 $[$rad$]$ & [pb/rad] & high  & low & high & low & high & low \\
\hline
 $[0.00,0.50)$ & $2.68$ & $+0.08$ & $-0.08$ & $+0.22$ & $-0.52$ & $+0.24$ & $-0.52$ \\
 $[0.50,1.00)$ & $3.10$ & $+0.09$ & $-0.09$ & $+0.25$ & $-0.36$ & $+0.26$ & $-0.37$ \\
 $[1.00,1.50)$ & $3.46$ & $+0.09$ & $-0.09$ & $+0.29$ & $-0.36$ & $+0.31$ & $-0.37$ \\
 $[1.50,1.75)$ & $4.51$ & $+0.15$ & $-0.15$ & $+0.36$ & $-0.42$ & $+0.39$ & $-0.44$ \\
 $[1.75,2.00)$ & $6.26$ & $+0.17$ & $-0.17$ & $+0.53$ & $-0.59$ & $+0.55$ & $-0.61$ \\
 $[2.00,2.25)$ & $8.93$ & $+0.20$ & $-0.20$ & $+0.73$ & $-1.02$ & $+0.76$ & $-1.04$ \\
 $[2.25,2.35)$ & $11.6$ & $+0.4$ & $-0.4$ & $+0.9$ & $-1.1$ & $+1.0$ & $-1.2$ \\
 $[2.35,2.45)$ & $13.9$ & $+0.4$ & $-0.4$ & $+1.1$ & $-1.3$ & $+1.1$ & $-1.4$ \\
 $[2.45,2.55)$ & $17.4$ & $+0.4$ & $-0.4$ & $+1.3$ & $-1.6$ & $+1.4$ & $-1.7$ \\
 $[2.55,2.65)$ & $21.8$ & $+0.5$ & $-0.5$ & $+2.0$ & $-2.2$ & $+2.0$ & $-2.3$ \\
 $[2.65,2.70)$ & $26.7$ & $+0.8$ & $-0.8$ & $+2.1$ & $-2.4$ & $+2.3$ & $-2.5$ \\
 $[2.70,2.75)$ & $29.6$ & $+0.8$ & $-0.8$ & $+3.2$ & $-3.4$ & $+3.3$ & $-3.5$ \\
 $[2.75,2.80)$ & $35.8$ & $+0.9$ & $-0.9$ & $+2.8$ & $-3.1$ & $+3.0$ & $-3.3$ \\
 $[2.80,2.85)$ & $42.9$ & $+1.0$ & $-1.0$ & $+3.3$ & $-3.8$ & $+3.4$ & $-3.9$ \\
 $[2.85,2.90)$ & $48.4$ & $+1.1$ & $-1.1$ & $+3.8$ & $-4.4$ & $+3.9$ & $-4.6$ \\
 $[2.90,2.95)$ & $57.4$ & $+1.2$ & $-1.2$ & $+4.3$ & $-4.6$ & $+4.4$ & $-4.7$ \\
 $[2.95,3.00)$ & $71.7$ & $+1.3$ & $-1.3$ & $+5.4$ & $-5.9$ & $+5.6$ & $-6.1$ \\
 $[3.00,3.05)$ & $80.8$ & $+1.4$ & $-1.4$ & $+6.1$ & $-7.4$ & $+6.3$ & $-7.6$ \\
 $[3.05,3.10)$ & $100.5$ & $+1.6$ & $-1.6$ & $+7.3$ & $-8.5$ & $+7.4$ & $-8.6$ \\
 $[3.10,3.14)$ & $107.6$ & $+1.8$ & $-1.8$ & $+7.9$ & $-9.6$ & $+8.1$ & $-9.8$ \\
\hline \hline
\end{tabular}
    \caption{
      Experimental cross-section values per bin in pb/rad for \dphigg.
      The listed total errors are the quadratic sum of statistical and systematic uncertainties.
    }
    \label{tab:cross_section_dphigg}
  \end{center}
\end{table}

\begin{table}[!hb]
  \begin{center}
\begin{tabular}{lrrrrrrr}
\hline \hline
 \cosThetagg  & d$\sigma$/d\cosThetagg & \multicolumn{2}{c}{Statistical error} & \multicolumn{2}{c}{Systematic errors} & \multicolumn{2}{c}{Total error} \\
  & [pb] & high  & low & high & low & high & low \\
\hline
 $[-1.00,-0.92)$ & $11.0$ & $+0.4$ & $-0.4$ & $+1.2$ & $-1.6$ & $+1.3$ & $-1.7$ \\
 $[-0.92,-0.84)$ & $14.5$ & $+0.4$ & $-0.4$ & $+1.1$ & $-1.6$ & $+1.2$ & $-1.6$ \\
 $[-0.84,-0.76)$ & $16.5$ & $+0.5$ & $-0.5$ & $+1.3$ & $-1.6$ & $+1.4$ & $-1.6$ \\
 $[-0.76,-0.68)$ & $18.8$ & $+0.5$ & $-0.5$ & $+1.4$ & $-1.6$ & $+1.5$ & $-1.7$ \\
 $[-0.68,-0.60)$ & $21.8$ & $+0.6$ & $-0.6$ & $+1.5$ & $-1.9$ & $+1.6$ & $-1.9$ \\
 $[-0.60,-0.52)$ & $22.7$ & $+0.6$ & $-0.6$ & $+1.8$ & $-2.0$ & $+1.9$ & $-2.1$ \\
 $[-0.52,-0.44)$ & $23.9$ & $+0.6$ & $-0.6$ & $+1.8$ & $-2.1$ & $+1.9$ & $-2.1$ \\
 $[-0.44,-0.36)$ & $24.6$ & $+0.6$ & $-0.6$ & $+1.8$ & $-2.0$ & $+1.9$ & $-2.1$ \\
 $[-0.36,-0.28)$ & $25.2$ & $+0.6$ & $-0.6$ & $+1.8$ & $-2.2$ & $+1.9$ & $-2.3$ \\
 $[-0.28,-0.20)$ & $28.3$ & $+0.6$ & $-0.6$ & $+2.1$ & $-2.6$ & $+2.2$ & $-2.7$ \\
 $[-0.20,-0.12)$ & $28.1$ & $+0.7$ & $-0.7$ & $+2.0$ & $-2.6$ & $+2.1$ & $-2.6$ \\
 $[-0.12,-0.04)$ & $29.6$ & $+0.7$ & $-0.7$ & $+2.3$ & $-2.7$ & $+2.4$ & $-2.8$ \\
 $[-0.04,0.04)$ & $31.4$ & $+0.7$ & $-0.7$ & $+2.5$ & $-2.9$ & $+2.6$ & $-3.0$ \\
 $[0.04,0.12)$ & $29.0$ & $+0.7$ & $-0.7$ & $+2.4$ & $-2.5$ & $+2.5$ & $-2.6$ \\
 $[0.12,0.20)$ & $27.8$ & $+0.7$ & $-0.7$ & $+2.1$ & $-3.0$ & $+2.2$ & $-3.1$ \\
 $[0.20,0.28)$ & $26.2$ & $+0.6$ & $-0.6$ & $+1.8$ & $-2.3$ & $+2.0$ & $-2.3$ \\
 $[0.28,0.36)$ & $25.7$ & $+0.6$ & $-0.6$ & $+2.0$ & $-2.1$ & $+2.1$ & $-2.2$ \\
 $[0.36,0.44)$ & $25.0$ & $+0.6$ & $-0.6$ & $+1.9$ & $-1.9$ & $+2.0$ & $-2.0$ \\
 $[0.44,0.52)$ & $24.9$ & $+0.6$ & $-0.6$ & $+1.7$ & $-2.0$ & $+1.8$ & $-2.1$ \\
 $[0.52,0.60)$ & $22.7$ & $+0.6$ & $-0.6$ & $+1.9$ & $-2.2$ & $+2.0$ & $-2.2$ \\
 $[0.60,0.68)$ & $21.3$ & $+0.6$ & $-0.6$ & $+1.6$ & $-1.8$ & $+1.7$ & $-1.9$ \\
 $[0.68,0.76)$ & $19.5$ & $+0.5$ & $-0.5$ & $+1.4$ & $-1.8$ & $+1.5$ & $-1.9$ \\
 $[0.76,0.84)$ & $16.9$ & $+0.5$ & $-0.5$ & $+1.4$ & $-1.6$ & $+1.5$ & $-1.7$ \\
 $[0.84,0.92)$ & $13.9$ & $+0.4$ & $-0.4$ & $+1.1$ & $-1.7$ & $+1.2$ & $-1.8$ \\
 $[0.92,1.00)$ & $11.7$ & $+0.4$ & $-0.4$ & $+1.0$ & $-1.4$ & $+1.1$ & $-1.4$ \\
\hline \hline
\end{tabular}
    \caption{
      Experimental cross-section values per bin in pb for \cosThetagg.
      The listed total errors are the quadratic sum of statistical and systematic uncertainties.
    }
    \label{tab:cross_section_costh}
  \end{center}
\end{table}

\onecolumn 
\clearpage
\begin{flushleft}
{\Large The ATLAS Collaboration}

\bigskip

G.~Aad$^{\rm 48}$,
T.~Abajyan$^{\rm 21}$,
B.~Abbott$^{\rm 111}$,
J.~Abdallah$^{\rm 12}$,
S.~Abdel~Khalek$^{\rm 115}$,
A.A.~Abdelalim$^{\rm 49}$,
O.~Abdinov$^{\rm 11}$,
R.~Aben$^{\rm 105}$,
B.~Abi$^{\rm 112}$,
M.~Abolins$^{\rm 88}$
% ...
\bigskip
\\
% ...
$^{11}$ Institute of Physics, Azerbaijan Academy of Sciences, Baku, Azerbaijan\\
$^{12}$ Institut de F{\'\i}sica d'Altes Energies and Departament de F{\'\i}sica de la Universitat Aut{\`o}noma de Barcelona and ICREA, Barcelona, Spain\\
% ...
$^{21}$ Physikalisches Institut, University of Bonn, Bonn, Germany\\
% ...
$^{48}$ Fakult{\"a}t f{\"u}r Mathematik und Physik, Albert-Ludwigs-Universit{\"a}t, Freiburg, Germany\\
$^{49}$ Section de Physique, Universit{\'e} de Gen{\`e}ve, Geneva, Switzerland\\
% ...
$^{88}$ Department of Physics and Astronomy, Michigan State University, East Lansing MI, United States of America\\
% ...
$^{105}$ Nikhef National Institute for Subatomic Physics and University of Amsterdam, Amsterdam, Netherlands\\
% ...
$^{111}$ Homer L. Dodge Department of Physics and Astronomy, University of Oklahoma, Norman OK, United States of America\\
$^{112}$ Department of Physics, Oklahoma State University, Stillwater OK, United States of America\\
% ...
$^{115}$ LAL, Universit{\'e} Paris-Sud and CNRS/IN2P3, Orsay, France\\
% ...
\end{flushleft}


\end{document}